\documentclass[fleqn,usenatbib]{mnras}

\usepackage{newtxtext,newtxmath}

\usepackage[T1]{fontenc}

\DeclareRobustCommand{\VAN}[3]{#2}
\let\VANthebibliography\thebibliography
\def\thebibliography{\DeclareRobustCommand{\VAN}[3]{##3}\VANthebibliography}

\usepackage{graphicx}	
\usepackage{amsmath}	
\usepackage{booktabs}
\usepackage{tabularx}
\usepackage{float}

\newcommand{\dd}{\mathrm{d}}

\title[Atmospheric Temperature Sensitivity to CIA]{Small Collisions, Big Impact: The Sensitivity of Atmospheric Temperatures to Collision-Induced Absorption}

\author[Dahlb\"udding \textit{et al.}]{
David Dahlb\"udding$^{1,2,3}$\thanks{E-Mail: ddahlb@mpe.mpg.de},
Tommaso Grassi$^{2,3}$\thanks{Corresponding Author E-Mail: tgrassi@mpe.mpg.de},
Karan Molaverdikhani$^{1,3}$,
Giulia Roccetti$^{4,5}$\thanks{ESA Research Fellow},\newauthor
Barbara Ercolano$^{1,2,3}$,
and Paola Caselli$^{2,3}$
\\
$^{1}$Universitäts-Sternwarte, Fakultät für Physik, Ludwig-Maximilians-Universität München, Scheinerstr. 1, 81679 München, Germany; \\
$^{2}$Centre for Astrochemical Studies, Max-Planck-Institut für extraterrestrische Physik, Gießenbachstr. 1, 85749 Garching, Germany; \\
$^{3}$Exzellenzcluster ‘Origins’, Boltzmannstr. 2, 85748 Garching, Germany; \\
$^{4}$European Space Agency (ESA), European Space Astronomy Centre (ESAC), Camino Bajo del Castillo s/n, Villanueva de la Cañada, E-28692 Madrid, Spain;\\
$^{5}$INAF — Osservatorio Astrofisico di Arcetri, Largo Enrico Fermi 5, I-50125 Firenze, Italy
}

\date{Accepted 2026 October 15. Received 2026 October 14; in original form 2026 August 10}

\pubyear{\the\year}

\begin{document}
\label{firstpage}
\pagerange{\pageref{firstpage}--\pageref{lastpage}}

\maketitle

\begin{abstract}
Collision-induced absorption (CIA) data are relevant to determining atmospheric temperatures and, in turn, the habitability of a diverse set of planetary atmospheres. However, these data remain sparse across broad wavelength and temperature ranges. Nevertheless, multiple, sometimes disagreeing, sources of experimental and theoretical data can be available for the same CIA pair. In this paper, we assess the impact of these varying sources on atmospheric temperatures and the limitations that the available data coverage can introduce. In order to inform future measurements and calculations, we test additional cases with extrapolated wavelength and temperature ranges. We first introduce a measure of the importance of CIA opacities depending on pressure and temperature by calculating the Rosseland mean with and without each CIA pair's contribution. Secondly, we employ each CIA source in a radiative transfer model of idealised atmospheres which maximise the effect of the CIA pair of interest. Our findings highlight the uncertainties related to the current data. Temperature differences between CIA data sources range from $\sim$10\,K when sources overlap in wavelength and temperature coverage to >100\,K when spectral gaps or temperature limits differ. Spectral windows or narrow temperature coverage can also hinder the model's convergence of the temperature profile. This work aims to provide guidance on the selection of CIA opacity datasets for atmospheric modelling and to identify the wavelength and temperature regimes in which further laboratory measurements and theoretical calculations are most urgently needed. We publish all our findings in detail in the form of an accessible website.
\end{abstract}




\section{Introduction}

Collision-Induced Absorption (CIA) arises from the interaction of colliding molecules --- supermolecular complexes --- which induce a transient dipole moment in otherwise non-polar species \citep[e.g.][]{Frommhold:2006}. The absorption coefficient of these transitions scales with the collision frequency, and hence with the square of the gas density (or, more precisely, the product of the densities of the colliding partners). CIA thus becomes a dominant source of opacity only in high-pressure environments $\gtrsim 1$\,bar. Crucially, this mechanism allows symmetric molecules such as $\text{N}_2$, $\text{H}_2$, and $\text{CH}_4$ to act as potent greenhouse gases, filling spectral windows left open by molecular absorption.

This absorption mechanism is fundamental to our understanding of diverse planetary climates. On the Archean Earth, $\text{N}_2$-$\text{H}_2$ absorption has been invoked to maintain habitable temperatures despite the Faint Young Sun \citep{Kasting:2013, Wordsworth:2013}. Similarly, the warming of Early Mars to allow for liquid water is often attributed to reducing atmospheres stabilised by $\text{CO}_2$-$\text{H}_2$ and $\text{CO}_2$-$\text{CH}_4$ interactions \citep{Ramirez:2014, Wordsworth:2017, Turbet:2020}. In the outer solar system, $\text{N}_2$-$\text{N}_2$ and $\text{N}_2$-$\text{CH}_4$ CIAs are critical for the thermal structure of Titan \citep{deKok:2010, Johnson:2024}, while $\text{H}_2$-$\text{H}_2$ and $\text{H}_2$-He dominate the opacity of gas giants \citep{Stevenson:1999, Guillot:2005}.

Hence, CIA is now a standard component of most radiative transfer codes used for planetary atmospheres, like \textsc{HELIOS} \citet{Malik:2017, Malik:2019a}, \textsc{petitCODE} \citep{Molliere:2015} and \textsc{AGNI} \citep{Nicholls:2024, Nicholls:2025a, Nicholls:2025b}. Databases such as HITRAN provide centralised repositories for these opacity data \citep{Richard:2011, Karman:2019, Terragni:2025}, yet the sensitivity of atmospheric models to the choice of CIA dataset is rarely systematically quantified. This study addresses that gap by evaluating the sensitivity of the greenhouse effect to variations in CIA data.

Recent work has quantified the significant impact of spectroscopic assumptions on exoplanetary atmospheric models, including pressure-broadening parameters \citep{Hedges:2016} and line-wing cutoffs \citep{Gharib-Nezhad:2024}. A comparable systematic assessment for CIAs is less common, although targeted studies exist, most notably the work by \citet{Turbet:2020} quantifying the climatic impact of updated $\text{CO}_2$-$\text{H}_2$ and $\text{CO}_2$-$\text{CH}_4$ measurements for Early Mars. More recently, \citet{Chaverot:2025} provided updated $\chi$ factors, i.e., the correction multipliers applied to the far wings of theoretical molecular absorption line shapes, for $\text{CO}_2$-$\text{CO}_2$, $\text{CO}_2$-$\text{N}_2$, and $\text{CO}_2$-$\text{H}_2\text{O}$ based on HITRAN 2020 \citep{Gordon:2021}. They extend temperature coverage to 3000 K and release new correlated-k tables for terrestrial planet modelling. Such accurate far-wing treatments represent a major step forward in handling continuum absorption and modelling temperature structures. However, our study addresses the parallel challenge of evaluating the sensitivity of these models to the varying underlying CIA data sources.

This propagation of differences between various available CIA data sources into climate metrics such as surface temperatures ($T_{\text{surf}}$) remains generally poorly constrained for a broader range of atmospheres. As is evident from the example of Early Mars, switching just one CIA opacity source can already lead to differences of tens of Kelvin in surface temperatures. Such significant swings can be crucial in correctly assessing whether a planet is habitable. 

Therefore, we aim to assess the impact of employing different CIA data sources and their associated uncertainties on atmospheric temperature structures. By identifying the regions of temperature, pressure, and wavelength where data discrepancies are most relevant, we determine where improved laboratory measurements or theoretical calculations are most necessary. To isolate these effects, we utilise simplified atmosphere models composed of specific collision pairs.

Systems where CIA is especially important are tidally heated exomoons around gas giants at a significant distance from their host star, or even unbound and free-floating through the Galaxy. Such moons could potentially hold on to thick atmospheres and retain enough internal heat for liquid water to exist at the surface \citep{Avila:2021, Roccetti:2023}. These high-pressure, but relatively low-temperature environments are ideal test cases for CIAs. As previous studies have shown, $\text{H}_2$-$\text{H}_2$ CIA in stable hydrogen-dominated atmospheres could conserve habitable conditions for up to 4.3 Gyr \citep{MolLous:2022, Roccetti:2023, Dahlbüdding:2026}. We use our previously developed atmospheric model, combining the \textsc{HELIOS} radiative transfer code \citep{Malik:2017, Malik:2019a, Malik:2019b, Whittaker:2022} with an iterative adjustment of molecular abundances, to accurately model temperature profiles.

In Section 2, we describe the selection of collision pairs from the HITRAN database, their analysis and their use in radiative transfer calculations. Section 3 explores the most interesting results of our analysis. Their impact is discussed in further detail in Section 4, along with possible limitations.

\section{Methods}

This study focuses on the greenhouse effect of the atmospheres of rocky planets. Because of this, we do not intend to analyse all available CIA opacities, but instead preselect pairs containing the major atmospheric constituents: $\text{H}_2$, $\text{N}_2$, $\text{CH}_4$ and $\text{CO}_2$ \citep[e.g.,][]{Woitke:2021, Brachmann:2025}.

This means we exclude water, which only has narrow CIA features at relatively short wavelengths (e.g. around 1.7\,$\mu$m for $\text{CO}_2$-$\text{H}_2\text{O}$ and around 4\,$\mu$m for $\text{H}_2$-$\text{H}_2\text{O}$). It is usually handled separately in the water continuum \citep[e.g.][]{Mlawer:2023}, as it can be hard to disentangle the CIA component from the rest of the continuum contributions \citep{Vigasin:2014, Karman:2019}. Furthermore, we exclude all pairs containing noble gases. Because of their high volatility and lack of geological sources, they are unlikely to be a major component of any secondary, outgassed atmosphere \citep{Zahnle:1993}. Although helium-enriched or even helium-dominated atmospheres can remain after the selective hydrodynamic escape of a primordial envelope \citep{Hu:2015, Lammer:2025}, such scenarios are outside the scope of this work. Finally, we do not consider any pair that contains molecular oxygen. While their spectroscopic features could enable the detection of $\text{O}_2$ in exoplanets, they usually only cover a narrow wavelength range below $10\,\mu$m, where molecular absorption typically dominates, limiting their greenhouse effect.

The CIA pairs included in our analysis are listed in Table \ref{tab:cia_overview}. The source data are obtained from HITRAN\footnote{\url{https://hitran.org/cia/}}, with the main source indicated in bold. Only the $\text{H}_2$-$\text{H}_2$ CIA data from \citet{Borysow:2002} were downloaded separately.\footnote{\url{https://www.astro.ku.dk/~aborysow/programs/}} In the table, every source is also briefly commented on to motivate experiments, such as extrapolating the available data in wavelength or temperature space. All molecular opacities were downloaded from the DACE opacity database\footnote{\url{https://dace.unige.ch/opacityDatabase/}} \citep{Grimm:2021} and are listed in Appendix \ref{app:MolOpacs}, Table \ref{tab:opacities}.

For the CIA data to be usable in the radiative transfer code \textsc{HELIOS}, we have to convert them to the right format of a molecular opacity given in cm$^{2}$\,g$^{-1}$. The original HITRAN files are given in cm$^5$\,molecule$^{-2}$ or cm$^{-1}$\,(molecule cm$^{-3}$)$^{-2}$, i.e. absorption coefficient per number density squared, and only dependent on temperature. We hence linearly interpolate log(k [cm$^5$\,molecule$^{-2}$]) to the default \textsc{HELIOS} opacity grid with $\Delta T = 50$\,K. Using the ideal gas law, we can calculate the final opacity:

\begin{equation} \label{eq:opac_conversion}
    \alpha [\mathrm{cm}^{2}\,\mathrm{g}^{-1}] = \frac{P}{k_\mathrm{B} T} \frac{1}{\mu_2} k [\mathrm{cm}^{5}\,\mathrm{molecule}^{-2}],
\end{equation}
\\
with the Boltzmann constant $k_\mathrm{B}$, as well as temperature $T$ and pressure $P$ at the respective grid points (with intervals $\Delta (\log_{10} P / 1\,\text{bar}) = 1/3$). Following the \textsc{HELIOS} convention for CIA pairs, $\mu_2$ is the mean molecular weight of the second collision partner as listed in the pair name. During the radiative transfer calculations, these opacities are multiplied by a factor of $(\mu_2/\bar\mu)\,X_1 X_2$, with the total mean molecular weight $\bar\mu$ and mixing ratios $X_i$ of the collision pair, before being combined with other opacity sources. The factor $\mu_2$ therefore cancels exactly, and the resulting opacity of the mixture is independent of which partner is designated as the second one, as long as the same assignment is used in both steps.

We compared our own converted opacity files against the already available ones for $\text{H}_2$-$\text{H}_2$ CIA \citep{kitzmann:2025} for consistency. These supplied opacities seem to consist of multiple original datasets, as we found minimal differences with either \citet{Borysow:2002} (< 30\,$\mu$m) or \citet{Abel:2011} (> 30\,$\mu$m), depending on the wavelength. To avoid infinite values during the logarithmic interpolation, we set the floor value to $10^{-15}$\,cm$^{2}$\,g$^{-1}$, equal to that of the molecular opacities from the DACE opacity database.

\begin{table*}
\centering
\caption{Overview Table of CIA Sources and Modifications}
\label{tab:cia_overview}
\begin{tabularx}{\textwidth}{l l l X}
\toprule
\textbf{Pair} & \begin{tabular}[c]{@{}l@{}}\textbf{HITRAN}\\ \textbf{Name}\end{tabular} & \textbf{Source} & \textbf{Comments \& Further Test Cases} \\
\midrule
CH$_4$-CH$_4$ & \textbf{2011} & \citet{Borysow:Frommhold:1987}  & \begin{tabular}[c]{@{}l@{}}Extrapolate down to 9\,$\mu$m  to close $\sim$10\,$\mu$m gap\\ Additionally extrapolate to 600\,K\end{tabular} \\
\midrule
CO$_2$-CH$_4$ & \begin{tabular}[c]{@{}l@{}}2024 (Alt)\\ 2024 (Alt)\\ \textbf{2024 (Main)}\end{tabular} & \begin{tabular}[c]{@{}l@{}}\citet{Wordsworth:2017}\\ \citet{Turbet:2020}\\ \citet{Fakhardji:2022}\end{tabular} & \begin{tabular}[c]{@{}l@{}}Extrapolate 2017 down to 150\,K for stability\\ Experimental data of 2020 lower $\rightarrow$ reduce by $\sim$40\,\% (1$\sigma$)\\ \phantom{x}\end{tabular}\\
\midrule
CO$_2$-CO$_2$ & \begin{tabular}[c]{@{}l@{}}2011\\ \textbf{2024}\end{tabular} & \begin{tabular}[c]{@{}l@{}}\citet{Gruszka:Borysow:1997}\\ \citet{Gruszka:Borysow:1997}$^a$\end{tabular} & \begin{tabular}[c]{@{}l@{}}Alt is part of Main but has narrower wavelength coverage $\ge$ 30\,$\mu$m\\ Clamp 7\,$\mu$m feature to 500\,K value, up to 800\,K\end{tabular}\\
\midrule
CO$_2$-H$_2$ & \begin{tabular}[c]{@{}l@{}}2018\\ \textbf{2024}\end{tabular} & \begin{tabular}[c]{@{}l@{}}\citet{Wordsworth:2017}\\ \citet{Turbet:2020}\end{tabular} & \begin{tabular}[c]{@{}l@{}}Extrapolate 2017 down to 150\,K for stability\\ Vary 2020 by $\sim$20\,\% (1$\sigma$)\end{tabular}\\
\midrule
H$_2$-CH$_4$ & \textbf{eq\_2011} & \citet{Borysow:Frommhold:1986a} & $\pm15\,\% (1\sigma)$; extrapolate to 600\,K \\
\midrule
H$_2$-H$_2$ & \begin{tabular}[c]{@{}l@{}}N/A$^b$\\ \textbf{2011}\\ N/A$^b$ \& eq\_2018\\ 2011 \& eq\_2018\end{tabular} & \begin{tabular}[c]{@{}l@{}}\citet{Borysow:2002}\\ \citet{Abel:2011}\\ \citet{Borysow:2002}$^c$\\ \citet{Abel:2011}$^c$\end{tabular} & \\
\midrule
N$_2$-CH$_4$ & \begin{tabular}[c]{@{}l@{}}2011\\ \textbf{2024}\end{tabular} & \begin{tabular}[c]{@{}l@{}}\citet{Borysow:Tang:1993} \\ \citet{Finenko:2022} \end{tabular} & \begin{tabular}[c]{@{}l@{}} \phantom{x}\\ Extrapolate down to 50\,K or use constant 70\,K data point \end{tabular}\\
\midrule
N$_2$-H$_2$ & \textbf{2024} & \citet{Borysow:Frommhold:1986b}$^d$ & Extrapolate to 600\,K \\
\midrule
N$_2$-N$_2$ & \begin{tabular}[c]{@{}l@{}} LowT\_2018\\ \textbf{2021}\\ \phantom{x} \end{tabular} & \begin{tabular}[c]{@{}l@{}}\citet{Sung:2016}\\ \citet{Lafferty:1996}$^e$\\ \phantom{x} \end{tabular} & \begin{tabular}[c]{@{}l@{}} \phantom{x}\\ Max. +10\,\% at high $\lambda$ (as reported in \citet{Karman:2019});\\ Extrapolate or clamp down to 50\,K. \end{tabular}\\
\bottomrule
\end{tabularx}

\vspace{0.2cm}
\raggedright
\footnotesize
$^a$ Also includes \citet{Baranov:2004}, \citet{Terragni:2025}, \citet{Baranov:2003}, and \citet{Tran:2024}.\\
$^b$ \citet{Borysow:2002} is not part of the HITRAN database.\\
$^c$ Low temperature ($\leq$ 400\,K) data overwritten with \citet{Fletcher:2018} at available wavelengths ($\lambda \gtrsim 18.5\,\mu$m).\\
$^d$ Also includes \citet{Wei:2024}.\\
$^e$ Also includes \citet{Baranov:2005}, \citet{Hartmann:2017}, and \citet{Chistikov:2019}.
\end{table*}

\subsection{Importance Maps} \label{sec:importance}

\begin{figure*}
    \centering
    \includegraphics[width=\textwidth]{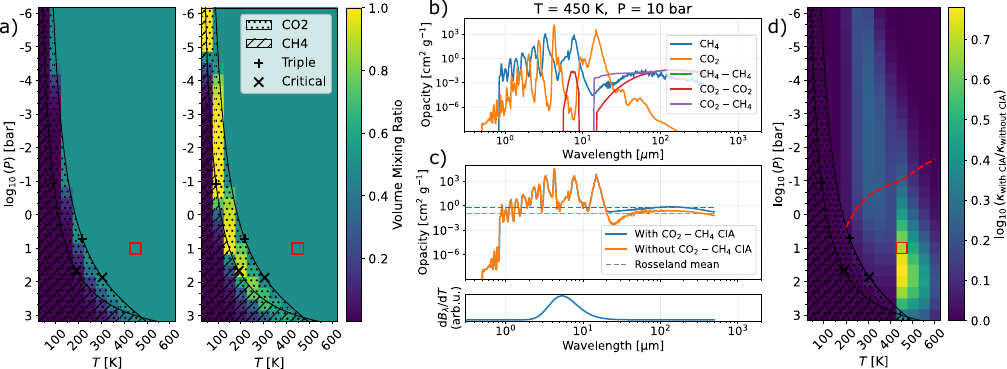}
    \caption{Importance Map Construction with the example of $\text{CO}_2$-$\text{CH}_4$ CIA: a) Volume mixing ratios dependent on pressure and temperature, with maximum VMR as determined by the saturation vapour pressure at a given temperature. Vapour pressure curves were extrapolated beyond the triple point (+) and critical point ($\times$). b) The spectral components at the given temperature and pressure ($P = 10$\,bar and $T = 450$\,K), consisting of the molecular absorption of $\text{CO}_2$ and $\text{CH}_4$, as well as the CIA contributions of $\text{CO}_2$-$\text{CO}_2$ and $\text{CO}_2$-$\text{CH}_4$ ($\text{CH}_4$-$\text{CH}_4$ on the other hand is at $10^{-15}$\,cm$^2$g$^{-1}$ because this lies beyond its maximum temperature of 400\,K). c) Combined opacities, one time with $\text{CO}_2$-$\text{CH}_4$ CIA (blue) and one time without (orange), with the respective Rosseland mean opacities shown in dashed lines. Below, the temperature derivative of the Planck function at 450\,K, which is used to calculate the Rosseland mean, is shown. d) The final importance map, which shows the logarithmic ratio of the Rosseland mean opacities with and without the $\text{CO}_2$-$\text{CH}_4$ CIA component and its dependence on pressure and temperature.}
    \label{fig:importance_map_contruction}
\end{figure*}

Before calculating the corresponding greenhouse effect produced by the CIA of each molecular pair, we assess at which pressure-temperature combinations a specific CIA is most relevant.

For this, we create a metric $I(P, T)$ represented as a map of importance for each pair and source to identify regions of interest: If these regions of high importance are near the limits of the temperature range of a source, this suggests to test the effects of extrapolating the data in temperature and spectral range, and gives relevant information to plan future calculations and measurements of CIA opacities. Furthermore, if a region shows particularly marked deviations between data sources, this suggests we should probe specific temperature ranges in more detail with our atmospheric model.

In the following, we will describe the calculation of this importance map using the example of $\text{CO}_2$-$\text{CH}_4$ CIA. First, we calculate the absorption spectrum at each $P,T$-combination, which can consist of up to five sources of opacity: two molecular ones (in our case $\text{CO}_2$ and $\text{CH}_4$) and three CIA contributions ($\text{CO}_2$-$\text{CO}_2$, $\text{CH}_4$-$\text{CH}_4$ and $\text{CO}_2$-$\text{CH}_4$), as seen in Fig.\,\ref{fig:importance_map_contruction}b. To maximise the absorption of the CIA, i.e. to maximise the collision frequency of the pair of interest, the standard number mixing ratio is set to 50\% each, if the pair consists of two separate molecules, or 100\%, if the molecule is colliding with itself. This is the case for the shown example spectrum at P = 10\,bar and T = 450\,K, marked by the red rectangle in Fig.\,\ref{fig:importance_map_contruction}a.

At low temperatures, this may not accurately reflect the physical reality due to condensation. We therefore adjust the maximum possible mixing ratio $X_\text{max}$ defined by the saturation vapour pressure: $X_\text{max} = P_\text{sat}(T) / P$. If possible, the non-condensing collision partner fills up the remaining partial pressure. For example, at $P = 1$\,bar and $T = 150$\,K, the saturation vapour pressure of $\text{CO}_2$ at this temperature would be around 0.06\,bar, resulting in a maximum mixing ratio of 6\,\%. Since the given total pressure is still below the saturation vapour pressure of $\text{CH}_4$ its mixing ratio is adjusted to 94\,\%. If, on the other hand, both molecules condense, then the remaining part is occupied by an invisible absorber with a mean molecular weight $\mu$ equal to that of the last condensing molecule. While this $\mu$ is not relevant to the value of the importance, it feeds into the scale height of the atmosphere, which will become relevant later on. The resulting mixing ratios are shown for our example of $\text{CO}_2$-$\text{CH}_4$ in Fig.\,\ref{fig:importance_map_contruction}a.

To reduce the intricate absorption spectra at each $P, T$ grid point to a single measure of importance, we employ the Rosseland mean opacity:

\begin{equation}
    \frac{1}{\kappa_{\text{R}}(P, T)} = \frac{\int_{0}^{\infty} \frac{1}{\kappa(P, T, \lambda)} \frac{\dd B(T, \lambda)}{\dd T} \dd\lambda}{\int_{0}^{\infty} \frac{\dd B(T, \lambda)}{\dd T} \dd\lambda}.
\end{equation}

It averages the inverse of the opacity over the temperature-derivative of the Planck function $B(T, \lambda)$ and is valid in the optically thick diffusion limit. By emphasising frequencies of low opacity, it accurately models the spectral windows in this regime \citep[see, e.g.,][]{Seager:2010, dePater:Lissauer:2015}. Since one of the most important factors in determining the greenhouse effect is the location of the radiative-convective boundary (RCB), namely, where the atmosphere becomes optically thick, the Rosseland mean is the appropriate choice for our purposes.

For each CIA pair and $P, T$ combination, we then calculate two Rosseland mean opacities: one with all spectral components, including the particular CIA contribution of interest, and the other without this specific CIA, as shown in Fig.\,\ref{fig:importance_map_contruction}c. The final importance metric is then the logarithmic ratio of these two opacities:

\begin{equation}
    I(P,T) = \log_{10}\left( \frac{\kappa_{\text{R, with CIA}}(T, P)}{\kappa_{\text{R, without CIA}}(T, P)} \right).
\end{equation}

To also illustrate the region of interest around the RCB in the importance plot, we estimate this boundary using $\kappa = \kappa_{\text{R, with CIA}}$ and assuming Earth-like gravity ($g$). Setting the optical depth over one scale height $H = k_\text{B}T / \mu g$ to 1, yields $P_\text{RCB}(T) = g \, \kappa^{-1}$, resulting in the red dashed line shown in Fig.\,\ref{fig:importance_map_contruction}d.

\subsection{Atmospheric Model}

\begin{figure}
\begin{center}
    \begin{minipage}{.24\textwidth}
        \includegraphics[width=\textwidth]{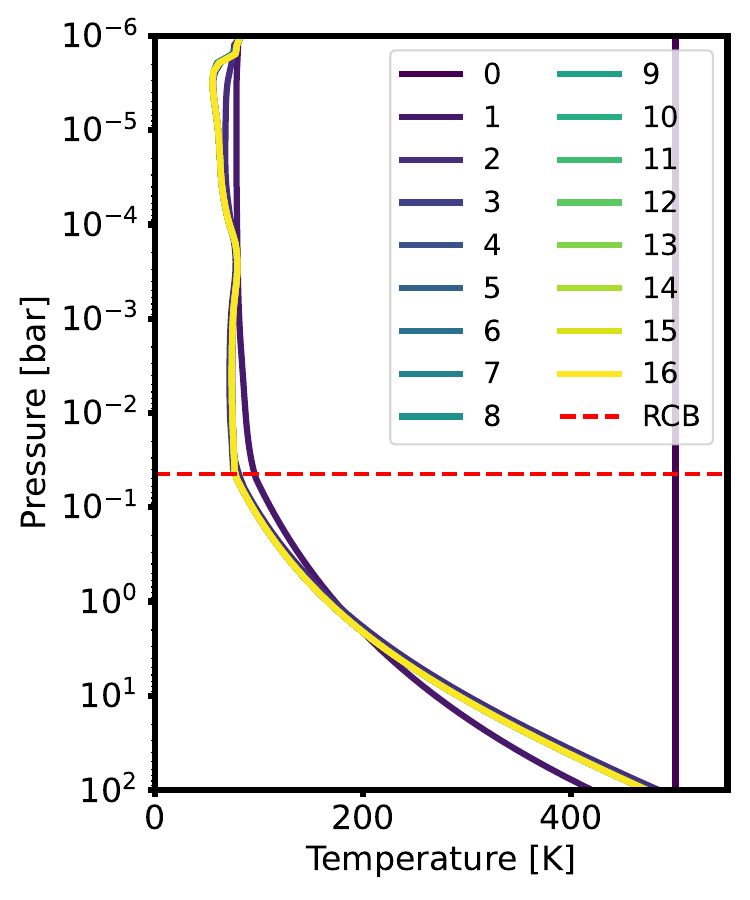}
    \end{minipage}%
    \begin{minipage}{.24\textwidth}
        \includegraphics[width=\textwidth]{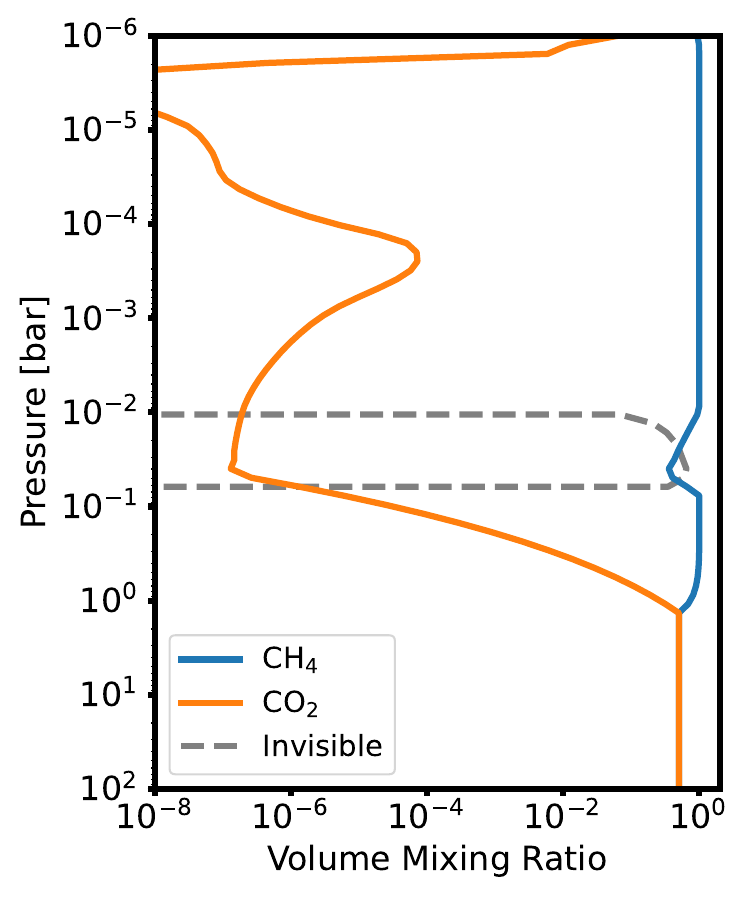}
    \end{minipage}%
    \caption{Example run of the coupled \textsc{HELIOS} code for the $\text{CO}_2$-$\text{CH}_4$ CIA pair with P$_\mathrm{surf}$ = 100 bar and T$_\mathrm{int}$ = 150 K. Left: Evolution of the pressure-temperature profile T(P) with the radiative-convective boundary (RCB) of the last step. Right: Final VMR profiles of all molecules.}
    \label{fig:iteration}
\end{center}
\end{figure}

The estimated location of the RCB, combined with the convective temperature gradient, could already allow us to estimate surface temperatures. But a full radiative-transfer model better captures the effects of small variations in opacity on the atmosphere's temperature structure. For this, we employ the open-source code \textsc{HELIOS}.

For each CIA data source, we run a range of models that vary only the effective temperature $T_\text{eff}$. Together with the greenhouse effect, it is the main factor determining surface temperatures. In the case of tidally heated exomoons around free-floating planets it is equal to the internal temperature, i.e. $T_\text{eff} = T_\text{int}$. The equilibrium temperature $T_\text{eq}$, as determined by the incident radiation and the albedo, on the other hand, does not contribute significantly.

For the atmospheres of such moons, we use the previously developed coupled \textsc{HELIOS} routine from \citet{Dahlbüdding:2026}. This time, the equilibrium chemistry of \textsc{GGchem} \citep{Woitke:2018} is replaced with an adjustment of the abundances of the collision partners, according to their respective vapour pressure curves, as described in Section \ref{sec:importance}. This means that, for each CIA pair, the atmosphere consists solely of the one or two molecules of interest, as shown in Fig.\,\ref{fig:iteration} for the example of $\text{CO}_2$-$\text{CH}_4$.

Although the total opacities would be predetermined, we continue to use an iterative scheme in which the temperature profile and condensation-restricted molecular abundances are updated repeatedly. This is because the native \textsc{HELIOS} grid, with a temperature step of $\Delta T = 50$ K, is too coarse to capture the rapid changes in mixing ratios along the condensation lines. We calculated that errors of a pre-computed grid, measured as the Rosseland mean opacity difference, exceeded one order of magnitude between grid points for nearly all example atmospheres. At the same time, the use of a finer grid is constrained by available memory.

We update our code in two ways: Firstly, \textsc{HELIOS}' on-the-fly opacity mixing now interpolates opacities in logarithmic $\log\kappa(\log P, T)$-space, instead of the previous linear interpolation ($\kappa(\log P, T)$). While this leads to sharp drops in opacity at the boundaries of the available data, which can hinder convergence, it improves accuracy within the valid range.

Secondly, we make use of the possibility to provide a grid of dry adiabatic lapse rates to \textsc{HELIOS}, which determines the logarithmic temperature-pressure gradient in the convective part of the atmosphere. The data for the specific heat capacity $c_\text{P}$ for each molecule to calculate the respective gradient are read from the \textsc{GGchem} files of the JANAF data \citep{Chase:1982, Chase:1986}.

We successfully validate our approach against the Early Mars model of \citet{Turbet:2020} by comparing the resulting surface temperatures, as described in more detail in Appendix \ref{app:EarlyMars}.

For all atmospheric models, we set the surface pressure to 100\,bar. Under these pressures, CIAs are highly relevant, and atmospheres are optically thick. At the same time, this allows us to compare temperatures at lower pressures without running additional models with different surface pressures, by interpolating the temperature profile $T(P, P_\text{surf}=100\,\text{bar})$ to the pressure of interest $P$. However, this interpolated temperature is only equal to the surface temperature of a model with lower surface pressure $T_\text{surf}(P_\text{surf}=P)$ if the atmosphere is still optically thick at pressure $P$, otherwise the interpolated temperature underestimates the respective surface temperature.

\subsection{Caveats}

Before interpreting the temperature deviations of the atmospheric model between different CIA data sources, several limitations inherent to our simplifications must be considered. Firstly, our assumption of binary atmospheres intentionally isolates the effects of specific CIA pairs, but this approach compromises some physical and chemical realism. For example, treating condensation as the only constraint on a molecule's mixing ratio can lead to an increase in the mean molecular weight in the upper radiative part of the atmosphere, which would be unstable according to \citet{Ledoux:1947}. 

Additionally, haze or cloud formation could diminish the relevance of CIA opacities. Especially low CIA opacity values, such as the ones extrapolated to lower wavelengths to close spectral gaps (e.g. $\text{CH}_4$-$\text{CH}_4$), could be overshadowed by these or other continuum opacity sources. 

Furthermore, the native pressure-temperature grid of \textsc{HELIOS} employs a temperature step of $\Delta T=50$\,K. While this is standard for many radiative transfer applications, it is relatively coarse compared to the grids on which some high-resolution CIA data are provided. This limits our ability to perfectly resolve rapid changes in opacity, especially near the lowest temperatures. 

Finally, the native behaviour of our script, which converts CIA data to the \textsc{HELIOS} format, sets these opacities to a floor value outside the available wavelength and temperature range. While this approach is agnostic to the actual behaviour outside those ranges, it is unphysical. Coupled with our updated logarithmic interpolation in \textsc{HELIOS}, this boundary treatment creates artificially sharp edges in the opacity profile, which directly contributes to the convergence issues discussed in further detail below. At the same time, if any CIA opacity is irrelevant beyond a certain temperature, this should not create any issues or differences in our models. Therefore, this sharp drop-off most clearly indicates the need to extend the data beyond the currently available range. While extending the boundaries via extrapolation can sometimes be useful to achieve numerical convergence, these extrapolated opacities can yield unphysical results. Hence, the corresponding temperature profiles must be interpreted with caution and, if physically implausible, potentially discarded.

\section{Results} \label{sec:results}

To evaluate the sensitivity of planetary climates to CIA data, we analysed both the P-T importance maps and the resulting temperature shifts in our coupled \textsc{HELIOS} models. A comprehensive overview of all collision pairs and plots is available online \footnote{\url{https://daviddahlbudding.github.io/CIA-Notebooks/}}. Here, we focus on the pairs that demonstrate the most significant climatic impacts or modelling challenges, divided into small (<50\,K) and large (>50\,K) temperature deviations, as well as differences in convergence behaviour within our setup.

\subsection{Small Temperature Deviations} \label{subsec:smallTdiffs}

\begin{figure*}
    \centering
    \includegraphics[width=\linewidth]{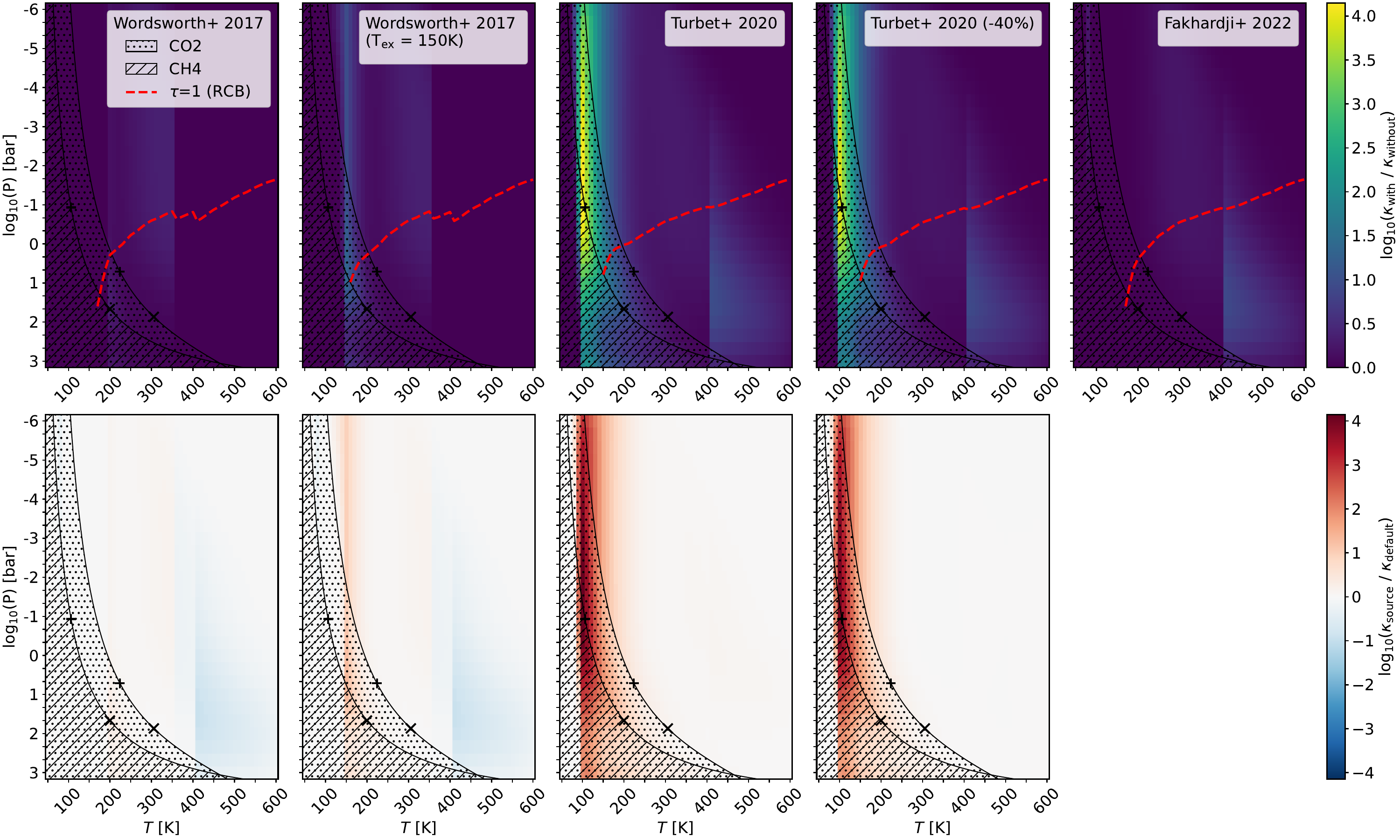}
    \caption{Importance map of $\text{CO}_2$-$\text{CH}_4$ CIA, with differences between the use of varying $\text{CO}_2$-$\text{CH}_4$ CIA opacity sources from \citet{Wordsworth:2017} (extrapolated from 200 to 150\,K), \citet{Turbet:2020} (reduced by 40\% in line with experimental results and uncertainty) and \citet{Fakhardji:2022} (default). The colorbar limits of the difference maps were adjusted to $\pm2$ dex for better visibility of smaller deviations, although higher values of up to 4.15\,dex were reached.}
    \label{fig:co2-ch4_importance}
\end{figure*}

\begin{figure}
    \centering
    \includegraphics[width=\linewidth, trim=0 1.5cm 1.5cm 2.5cm, clip]{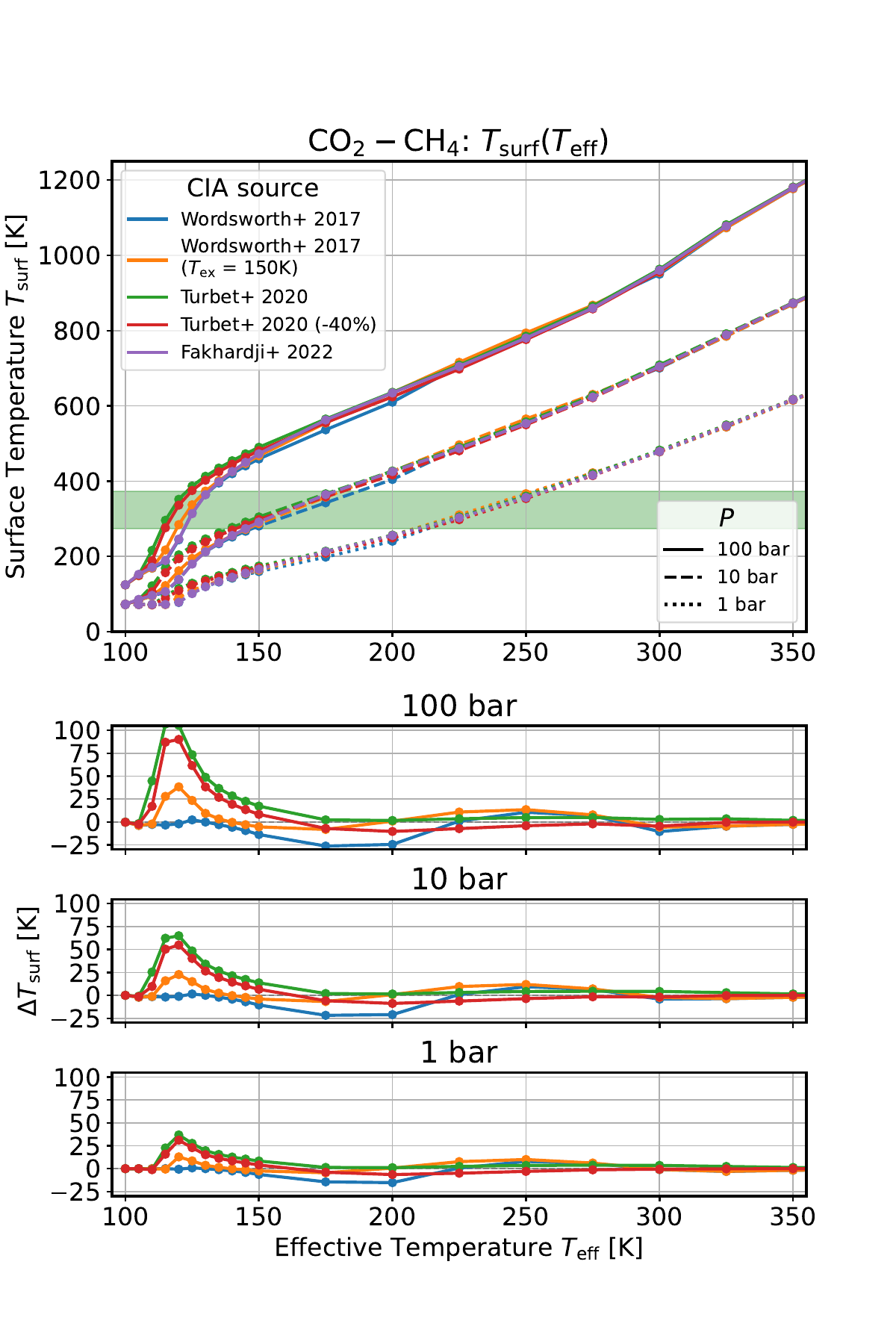}
    \caption{Surface temperatures of the binary atmosphere model consisting of $\text{CO}_2$ and $\text{CH}_4$, with differences between the use of varying $\text{CO}_2$-$\text{CH}_4$ CIA opacity sources from \citet{Wordsworth:2017} (extrapolated from 200 to 150\,K), \citet{Turbet:2020} (reduced by 40\% in line with experimental results and uncertainty) and \citet{Fakhardji:2022}. The green shaded area shows the habitable temperature range of 0-100\,$^\circ$C.}
    \label{fig:co2-ch4}
\end{figure}

For sources with similar wavelength and temperature coverage, we observe relatively small temperature deviations of up to a few tens of Kelvin.

\subsubsection{$\text{CO}_2$-$\text{CH}_4$}

A relevant example of this is $\text{CO}_2$-$\text{CH}_4$, whose importance maps are shown in Fig.\,\ref{fig:co2-ch4_importance} and atmospheric temperatures in Fig.\,\ref{fig:co2-ch4}. Above $T_\text{eff} = 150$\,K, atmospheric temperatures deviate by less than 15\,K, even at 100\,bar for all but the \citet{Wordsworth:2017} data, which shows lower surface temperatures by -25\,K versus the default \citep{Fakhardji:2022} source. The \citet{Turbet:2020} source yields higher or equal temperatures across all probed effective temperatures, with an increase of +5\,K spanning 250 - 300\,K. These small temperature differences correlate with the Rosseland mean differences shown in the third column of the second row of Fig.\,\ref{fig:co2-ch4_importance}. For temperatures above 230\,K, the Rosseland mean calculated with \citet{Turbet:2020} CIA data slightly exceeds the default source \citep{Fakhardji:2022}, but by less than a factor of 2.

We also reduced the opacities of \citet{Turbet:2020} to 60\,\% of their original values, in line with the measurement uncertainty of \citet{Turbet:2020} and the fact that their measurements appear consistently below the final semi-empirical model. This reduction results in around -5 to -10\,K lower surface temperatures in the 150\,K to 300\,K range of internal temperatures. While these opacity values are closer to the ones from \citet{Fakhardji:2022}, temperatures can be lower by -10\,K at $T_\text{eff} = 200$\,K, with differences vanishing above 300\,K.

The most significant differences in both mean opacities and temperatures in the 100\,K to 150\,K range will be addressed in the next section (\ref{subsec:large_Tdiffs}).

\subsubsection{Others}

Small surface temperature deviations can also be seen between the two sources of $\text{CO}_2$-$\text{H}_2$ (up to +14\,K at $T_\text{eff} = 300$\,K).\\

Varying the $\text{H}_2$-$\text{CH}_4$ opacity with the typical standard deviation for the underlying measurement of 15\% results in a maximum $\pm 7$\,K surface temperature variation. Extrapolating the temperature range from 400\,K to 600\,K, yields significantly higher temperatures (+34\,K) at $T_\text{eff} = 400$\,K.\\

Finally, overwriting the low temperature ($\leq$ 400\,K), long wavelength ($\lambda \gtrsim 18.5\,\mu$m) data of the \citet{Borysow:2002} $\text{H}_2$-$\text{H}_2$ CIA with the newest models from \citet{Fletcher:2018} leads to a fluctuating impact of -16\,K at $T_\text{eff} = 50$\,K to +10\,K in the $T_\text{eff} = [150-200]$\,K range, whereas overwriting the \citet{Abel:2011} data similarly leads to differences between -48\,K to +15\,K in the same $T_\text{eff}$-ranges.

\subsection{Large Temperature Deviations} \label{subsec:large_Tdiffs}

\begin{figure}
    \centering
    \includegraphics[width=\linewidth, trim=0 1.5cm 1.5cm 2.5cm, clip]{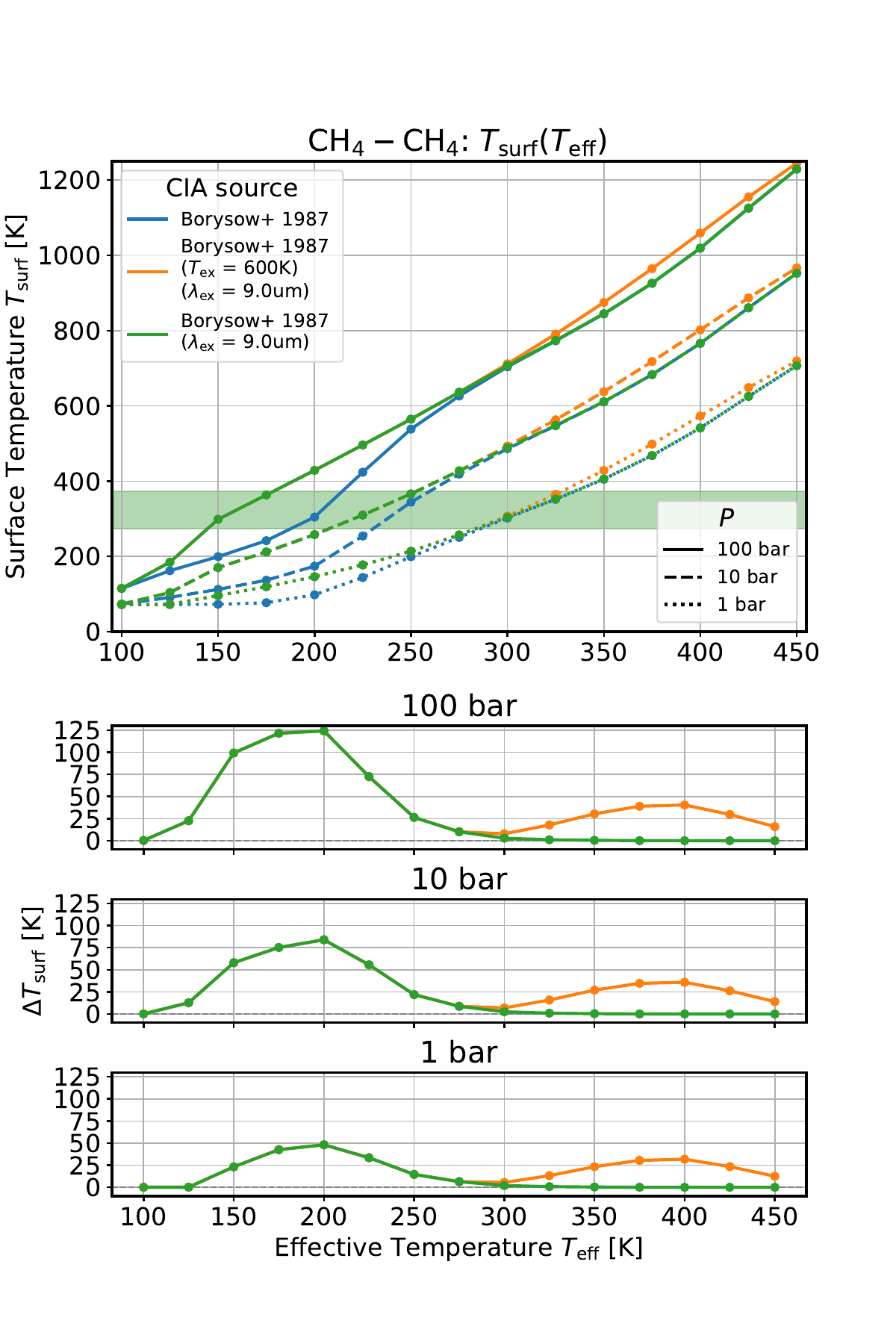}
    \caption{Surface temperatures of the atmosphere model consisting only of $\text{CH}_4$, with differences between the use of varying $\text{CH}_4$-$\text{CH}_4$ CIA opacities from \citet{Borysow:Frommhold:1987} (extrapolated down to 9\,$\mu$m and additionally from 400 up to 600\,K). The green shaded area shows the habitable temperature range of 0-100\,$^\circ$C.}
    \label{fig:ch4-ch4}
\end{figure}

Differences in the wavelength and temperature ranges of the available data can often lead to large deviations in atmospheric temperatures, as illustrated in Fig.\,\ref{fig:ch4-ch4} for the example of $\text{CH}_4$-$\text{CH}_4$. But even sources which broadly overlap in their available data ranges can have significantly different temperatures through feedback loops, as we will see in the following.

\subsubsection{$\text{CO}_2$-$\text{CH}_4$}

Most evident in the temperature difference plots of $\text{CO}_2$-$\text{CH}_4$ (right side of Fig.\,\ref{fig:co2-ch4}) for both the original and reduced \citet{Turbet:2020} data (the green and red lines, respectively) is the peak around $T_\text{eff} = 120$\,K, deviating by more than +100\,K at 100\,bar from temperatures obtained with the \citet{Fakhardji:2022} CIA data.

This is especially relevant considering the absolute surface temperatures (left side of the same Figure). For both 10 and 100\,bar, these positive temperature shifts result in a shift of the habitable zone as it depends on the effective temperature, i.e. the interval on the x-axis where a line crosses the green $T_\text{surf} = 0-100\,^\circ$C range. This, in turn, means that less internal heat is necessary to sustain habitable conditions.

This also overlaps with the dark red region in the difference maps (second row in Fig.\,\ref{fig:co2-ch4_importance}) between 100 and 150\,K. In addition to the opacity differences, the atmospheric models show that, at $T_\text{eff} = 150$\,K, $\text{CH}_4$ begins to condense near the radiative-convective boundary (see Fig.\,\ref{fig:iteration}), thereby exacerbating any temperature differences through a feedback loop between lower temperatures, more condensation, and lower optical depth.

\subsubsection{$\text{CH}_4$-$\text{CH}_4$}

Although only one source \citep{Borysow:Frommhold:1987} is available for $\text{CH}_4$-$\text{CH}_4$, we ran two additional test cases: extrapolating to lower wavelengths to close the spectral gap at around 10\,$\mu$m and additionally extrapolating to higher temperatures from 400 to 600\,K. This demonstrates the relevance of such low-wavelength, high-temperature data. In the extreme case of a pure $\text{CH}_4$, the extrapolation in wavelength space not only leads to faster convergence, but also a surface temperature increase of more than +100\,K in the 150 and 200\,K range of internal temperatures. It should be noted that the power-law extrapolation to lower wavelengths likely overestimates the opacity relative to the expected exponential drop-off in wavenumber space. The extrapolation in temperature space leads to an additional warming of +30 to +40\,K in the $T_\text{eff} = [350-425]$\,K range. 

\subsubsection{$\text{CO}_2$-$\text{CO}_2$}

The data for $\text{CO}_2$-$\text{CO}_2$ CIA's 7\,$\mu$m feature, located directly in a spectral window of $\text{CO}_2$'s molecular absorption, are available up to 500\,K. Our importance maps indicate that it is still relevant, with values of about 1.5\,dex at 500\,K, only slowly decreasing with temperature.

Extending its range to 800\,K by assuming the 500\,K data remain constant can also increase surface temperatures by more than +100\,K for internal temperatures exceeding 400\,K, in agreement with the high values of the importance map.

\subsubsection{$\text{H}_2$-$\text{H}_2$}

Lastly, for $\text{H}_2$-$\text{H}_2$ CIA, the difference between \citet{Borysow:2002} and \citet{Abel:2011} source (with or without the \citet{Fletcher:2018} data) can similarly top 100\,K for the highest probed internal temperatures of 400 and 450\,K, with the \citet{Borysow:2002} data resulting in higher temperatures.

\subsection{Model Convergence}

The two previously shown examples of $\text{CO}_2$-$\text{CH}_4$ (Fig.\,\ref{fig:co2-ch4}) and $\text{CH}_4$-$\text{CH}_4$ (Fig.\,\ref{fig:ch4-ch4}) already illustrate the improvements in convergence of the temperature profile in \textsc{HELIOS}, and thereby our iterative scheme, for extended wavelength and temperature ranges. For each, the sources with limited temperature or wavelength coverage required relaxation of the temperature and VMR profiles to achieve convergence. Additionally, \textsc{HELIOS} had to be forced to enable convection, as it would otherwise remain stuck in the radiative loop until it reached the maximum number of iterations.

In the case of $\text{CO}_2$-$\text{CH}_4$, the oldest available source, \citet{Wordsworth:2017}, converges in the $T_\text{eff} = [200-300]$\,K range more easily if opacities are extrapolated to significantly below the condensation limit of $\text{CO}_2$ where the CIA contribution vanishes. Although the low upper limit of the available data, 350\,K, can still lead to instabilities.

For $\text{CH}_4$-$\text{CH}_4$, the temperature range is less of a problem than the spectral gap at 10\,$\mu$m between the molecular and CIA contribution. Closing it improves convergence with our setup, especially at lower temperatures, at or below internal temperatures of 250\,K.

\subsection{N$_2$ Pairs}

The CIA pairs involving nitrogen $\text{N}_2$-$\text{X}$, including $\text{X}=\text{CH}_4, \text{H}_2, \text{N}_2$, do not show any differences between sources or any further test cases.

For two of the three pairs, this is in agreement with the importance maps. The available data range of $\text{N}_2$-$\text{H}_2$ CIA up to 400\,K is already sufficient, as its importance peaks at a value of 0.6\,dex at 50 to 150\,K, depending on the pressure, and then drops off to below 0.1\,dex above $T_\text{eff} \geq 230$\,K. Hence, extrapolating the available data to 600\,K does not significantly affect the importance or the atmospheric temperatures. Similarly, $\text{N}_2$-$\text{N}_2$ CIA is only important at lower temperatures, because its major absorption feature is at even higher wavelengths than for $\text{N}_2$-$\text{H}_2$. This sparse opacity fails to make the atmosphere optically thick, even at high pressures, thereby destabilising the atmospheric model.

The remaining case of $\text{N}_2$-$\text{CH}_4$ does show a significant difference in importance between the older \citet{Borysow:Tang:1993} and the newer \citet{Finenko:2022} source, where the older one has a significantly higher Rosseland mean opacity from around 120 to 350\,K. Still, atmospheric temperatures between them vary by less than 0.1\,K.

\section{Discussion}

The results presented in Section \ref{sec:results} demonstrate that the sensitivity of atmospheric temperatures to CIA data is driven primarily by the completeness of their wavelength and temperature coverage rather than differences in absolute opacity magnitude. While variations between overlapping CIA datasets typically lead to modest temperature shifts of order 10–20 K, missing spectral regions or truncated temperature ranges can produce much larger deviations, in some cases exceeding 100 K. These effects arise from the role of CIA in filling spectral windows and controlling radiative leakage, particularly at wavelengths near the peak of the Planck function. In addition, our results show that the treatment of opacities outside their tabulated ranges can introduce unphysical numerical artefacts, such as convergence failures or artificial cooling. Both aspects are critical for robust predictions of atmospheric structure and climate.

However, we note that the tests shown here are not valid for every rocky planet, and that the impact of CIA opacities will vary depending on specific atmospheric conditions. We are particularly interested in internally heated exomoons, as they rely heavily on broad infrared CIA features to retain heat. The choice of a sparse, binary atmosphere, on the other hand, was made purely to isolate and maximise the effect of the specific CIA pairs. Therefore, the resulting temperature variations serve as limiting cases rather than universal outcomes.

\subsection{Convergence Issues}

In radiative transfer modelling, non-convergence of the temperature profile can sometimes be a physical consequence of the atmosphere lacking the necessary opacity to constrain  temperatures. This is evident in the $\text{N}_2$-$\text{N}_2$ models, where the extremely sparse CIA opacities in the absence of any molecular absorption cannot establish a stable thermal structure even at pressures of tens of bars. 

However, convergence failures are frequently numerical artefacts. As noted above, when CIA data limits are exceeded, the opacity drops to zero. Because we interpolate the opacities in logarithmic space, this creates a sharp, cliff-like boundary in the absorption spectrum. If these cut-offs occur at high temperatures where long-wavelength CIA is relatively irrelevant, the model remains stable. However, if the boundary truncates a feature near the peak of the Planck function, the solver struggles to converge. This underscores the danger of deploying limited-range CIA data without appropriate extrapolation schemes.

\subsection{Extension of $\lambda$- and T-range}

Our experiments on extrapolating available data demonstrate the critical importance of expanded wavelength and temperature ranges for model stability and accuracy. 

At the low-temperature boundary, pairs such as $\text{CO}_2$-$\text{CH}_4$ and $\text{CO}_2$-$\text{H}_2$ remain climatically relevant at temperatures slightly below 200\,K, where $\text{CO}_2$ begins to condense. The oldest dataset \citep{Wordsworth:2017}, meant for models of Early Mars, is only available to a minimum of 200\,K, which creates numerical problems. Extrapolating these opacities down to 150\,K successfully stabilises some of the models. Theoretically, a similar benefit applies to $\text{N}_2$ pairs, where the coldest available data points are often between 50 and 100\,K. While extrapolating these to 50\,K does not significantly alter the final temperatures in our binary models, it can improve convergence.

Similarly, extending wavelength coverage helps close spectral windows. In the case of $\text{CH}_4$-$\text{CH}_4$, closing the $\sim 10\,\mu$m gap between the CIA and molecular absorption prevents the artificial leakage of thermal radiation. In a more chemically diverse atmospheres, this gap could be mitigated by other molecular species, depending on their abundance: $\text{NH}_3$, for example, has an absorption feature around this wavelength and also $\text{H}_2\text{O}$ and $\text{CO}_2$ have significant cross sections with less of a drop-off towards longer wavelengths than $\text{CH}_4$.

For $\text{CO}_2$-$\text{CO}_2$, it is also critical to extrapolate the far-infrared feature from \citet{Gruszka:Borysow:1997} down to short wavelengths and extend the $7\,\mu$m feature to higher temperatures, where it is still relevant, if not overshadowed by e.g. $\text{H}_2\text{O}$, as also evident by the corresponding importance map.

Hence, extending these data ranges successfully prevents this unphysical radiation leakage, thereby not only improving the accuracy of the calculated temperature profile but also mitigating both convergence issues and extreme temperature artefacts, such as deep, detached radiative layers.

We want to emphasise here a fundamental difference in handling CIA opacity data with either a limited temperature or a limited wavelength range. In cases with narrow temperature limits, these opacities can be clamped to the boundary values. As shown in this paper, the correct order of magnitude of opacities yields only minor temperature differences. Limited wavelength coverage, however, is more difficult to handle, which is why the opacities are usually set to zero outside the available data range.

A relevant example of how even relatively low opacity values outside a given data range can be highly significant to atmospheric temperatures involves the two most recent sources of $\text{CO}_2$-$\text{CH}_4$: Already their importance maps reveal key differences in their wavelength coverage, with minimum wavelengths of 9\,$\mu$m \citep{Turbet:2020} and 14\,$\mu$m \citep{Fakhardji:2022}. This region is an important spectral gap between the $\text{CH}_4$'s 9\,$\mu$m and the $\text{CO}_2$'s 15\,$\mu$m feature, especially at low temperatures, where absorption lines become sharper and $\text{CO}_2$ begins to condense. In our atmospheric model, this high difference in importance translates to a 100\,K difference in surface temperatures.

\citet{Turbet:2020} similarly emphasize this spectral window below $\text{CO}_2$'s 15\,$\mu$m feature for $\text{CO}_2$-$\text{H}_2$ CIA. They note that cutting the CIA opacities above 550\,cm$^{-1}$ (below 18\,$\mu$m) decreases the surface temperature by 8\,K on Early Mars (for 1\,bar of CO$_2$ with 10\,\% H$_2$). Although in contrast to $\text{CO}_2$-$\text{CH}_4$, the $\text{CO}_2$-$\text{H}_2$ pair has some of its highest opacity in this truncated region.

\subsection{Greenhouse Sensitivity}

As demonstrated in Section \ref{subsec:smallTdiffs}, when different CIA sources largely agree in their spectral and thermal coverage, the resulting deviations in surface temperature are relatively small, typically on the order of 10 to 20\,K. This is the case for the two newer sources of $\text{CO}_2$-$\text{CH}_4$ and $\text{CO}_2$-$\text{H}_2$, as well as for $\text{H}_2$-$\text{CH}_4$ when varied with its measurement uncertainty.

However, in the context of planetary habitability, these small deviations are highly significant. A 10\,K shift induced purely by the choice of underlying opacity data can easily dictate whether a modelled planet maintains liquid water or undergoes global glaciation. For reference, changing Earth's equilibrium temperature by $\pm$10\,K would correspond to a roughly $\pm$16\,\% change in solar flux, or $\pm$8\,\% shift in its semi-major axis. Furthermore, in more comprehensive models, these initial temperature differences would likely be amplified by positive feedback loops, such as the ice-albedo feedback or the runaway greenhouse effect. Therefore, uncertainties in CIA data must be treated as a primary source of error in climate predictions.

Again, the $\text{CO}_2$-$\text{CH}_4$ case can serve as an example of how small deviations in opacity data can lead to extreme differences in atmospheric temperatures.

From the importance maps alone, we see that the highest values for the \citet{Turbet:2020} data, and therefore the highest differences to the default \citet{Fakhardji:2022} data, are located where $\text{CO}_2$ condenses. This demonstrates that, while the chosen 50:50 mixture maximises the CIA opacity, it does not necessarily maximise its importance relative to molecular absorption. While the $\text{CO}_2$'s contribution to the total opacity falls proportionally to its mixing ratio, the $\text{CO}_2$-$\text{CH}_4$ CIA component drops off relatively less as the remaining partial pressure is filled up with $\text{CH}_4$. For example, as we go from a 50:50 mixture to one of 10\,\% $\text{CO}_2$ and 90\,\% $\text{CH}_4$, the molecular absorption component is reduced by a factor of 5, whereas the mixed CIA component is only reduced by a factor of 2.8. As mentioned in the previous section, the CIA opacity is mainly obscured by the $\text{CO}_2$'s 15\,$\mu$m feature. Hence, the decreasing $\text{CO}_2$ abundance contributes to the observed spike in importance. Through this example, we want to emphasise that the calculated atmospheric temperature differences could be further enhanced with different mixing ratios rather than the 50:50 mixture used here.

In addition to the peak in importance in the 100 - 150\,K range, the atmospheric models show that the spike in atmospheric temperature differences around $T_\text{eff} = 125$\,K is exacerbated by methane condensing around the crucial radiative-convective boundary. This can significantly reduce optical depths and therefore cool the atmosphere, condensing even more $\text{CH}_4$. This feedback loop gives rise to the steepest part of the $T_\text{surf}(T_\text{eff})$ curve, shown in Fig.\,\ref{fig:co2-ch4}. Similar self-reinforcing cycles could also occur with other important greenhouse gases that are not necessarily part of the respective CIA pair, like the classical runaway greenhouse \citep{Kasting:1988, Nakajima:1992, Goldblatt:2013}.

\section{Conclusions}

In this paper, we analysed the available data for all CIA pairs that include $\text{H}_2$, $\text{N}_2$, $\text{CH}_4$ or $\text{CO}_2$. We first introduced a measure of importance, the logarithmic ratio of the Rosseland mean opacities with and without the respective CIA, to find temperatures and pressures of interest. Such regions in the P-T plane include areas of high importance for any one CIA data source, large differences in importance between sources, or significant importance values at the border of the available temperature range for the respective data source. From this, we could construct further tests of opacities with artificially extended data coverage, e.g., by extrapolating or clamping values outside their temperature range.

We then tested all available and extended sources in a simplified atmospheric model of an internally heated Earth-sized rocky body. The thick 100\,bar atmosphere consisted purely of the molecules of the CIA pair of interest. Although this limited number of absorption sources is not the most realistic configuration, it illustrates the maximum impact CIA opacities can have on atmospheric temperatures and clearly shows the temperature ranges through which they are relevant. Therefore, the resulting temperature variations serve, on the one hand, as an upper limit that the underlying CIA data can introduce. On the other hand, such variations could be exacerbated by feedback loops, such as the runaway greenhouse effect. Our own model illustrates such a feedback loop in the example of $\text{CO}_2$-$\text{CH}_4$, where methane began to condense around the radiative-convective boundary, lowering the opacity, further decreasing temperatures, and hence triggering a runaway cooling effect. This ultimately resulted in differences of hundreds of Kelvin in atmospheric temperatures between the available data sources.

In general, we found that the importance maps and their differences between CIA data sources often aligned with the observed atmospheric temperature changes. These could typically range from no changes, as for all $\text{N}_2$-containing pairs, to tens of Kelvin in the case of, e.g., $\text{CO}_2$-$\text{H}_2$ or $\text{H}_2$-$\text{CH}_4$. Even larger differences of hundreds of Kelvin occurred in only a few cases apart from $\text{CO}_2$-$\text{CH}_4$, where limited wavelength or temperature ranges allowed radiation to escape through large spectral windows, as was the case for $\text{CH}_4$-$\text{CH}_4$, $\text{CO}_2$-$\text{CO}_2$ and $\text{H}_2$-$\text{H}_2$. In some cases, these windows combined with narrow temperature ranges could even hinder the atmospheric model from converging.

These extreme differences in temperature or convergence behaviour point to a lack of data in the respective wavelength and temperature ranges. Some of these discrepancies in CIA data could be mitigated in more realistic atmospheres by other absorption sources. For example, the relatively low extrapolated opacity values of $\text{CH}_4$-$\text{CH}_4$ CIA could be eclipsed by $\text{NH}_3$'s molecular absorption or by other continuum opacity sources such as clouds or hazes.

In other cases, the observed temperature changes indicate that CIA remains clearly relevant at temperatures above the available range, especially for pairs containing $\text{CO}_2$. Smaller temperature changes, around $\pm 10$\,K, were observed for data that we varied with measurement uncertainties ranging from 15 to 40\,\%. This indicates that new measured or theoretically derived CIA data, even with relatively high uncertainties, could significantly improve the accuracy of the calculated atmospheric temperatures.

Until more comprehensive CIA data become available, our results support the following practical treatment: the CIA opacity should be set to zero, or to an appropriate opacity floor, outside the tabulated wavelength range, while the nearest tabulated boundary value should be adopted at temperatures outside the available range. This approach avoids potentially unreliable extrapolation beyond the domain covered by the underlying calculations or measurements.

For \textsc{HELIOS}, this includes some particularities. Because it only accepts molecular opacities, we have to convert the binary absorption coefficient to the right format. Hence, while the binary one can be constant with pressure and temperature, the final opacity is not, as evident from Eq.\,\ref{eq:opac_conversion}. Furthermore, it may make sense to extrapolate outside some temperature ranges due to \textsc{HELIOS}' coarse temperature grid. Because pairs like $\text{N}_2$-$\text{CH}_4$ and $\text{N}_2$-$\text{N}_2$ have data down to 70\,K and 78\,K, respectively, it is more accurate to extrapolate opacities down to \textsc{HELIOS}' minimum of 50\,K than to leave the data constant.

For ease of use, we make our script for converting HITRAN to \textsc{HELIOS} CIA opacities available to download (\url{https://doi.org/10.5281/zenodo.21873428}). Furthermore, the complete analysis suite, including importance maps and atmospheric models for all CIA pairs, is publicly accessible on a companion website for visualisation and download (\url{https://daviddahlbudding.github.io/CIA-Notebooks/}).These results provide a basis for assessing the choice of CIA dataset and the associated uncertainties in planetary-atmosphere models. They also highlight the need for further theoretical and experimental determinations of collision-induced absorption over broader temperature ranges and with improved accuracy.

\section*{Acknowledgements}

The authors would like to thank the anonymous reviewers for their constructive comments and valuable suggestions, which helped improve the quality of this manuscript.\\
This work was supported by the Deutsche Forschungsgemeinschaft (DFG, German Research Foundation), TRR 392 (CRC 392) "Molecular Evolution in Prebiotic Environments" - 521256690, Research Unit “Transition discs” - 325594231 and the Excellence Cluster ORIGINS funded by Germany’s Excellence Strategy EXC-2094 - 390783311. T.G. and P.C. acknowledge the financial support of the Max Planck Society. G.R. acknowledges support through the European Space Agency (ESA) Research Fellowship in Space Science.\\
The text of this paper has been refined with Grammarly, while parts of the code and the analysis website have been written using Copilot in VS Code and the Gemini chat interface.\\
This research has made use of the NASA Astrophysics Data System, and the Python packages \textsc{NumPy}, \textsc{SciPy} and \textsc{Matplotlib}.

\section*{Data Availability}

The analysis of all investigated CIA pairs is available here: \url{https://daviddahlbudding.github.io/CIA-Notebooks/}.\\
The script to convert CIA opacities from the HITRAN to the \textsc{HELIOS} format, as well as all the produced data is available on Zenodo: \url{https://doi.org/10.5281/zenodo.21873428}.\\
The code, which couples \textsc{HELIOS} to the condensation adjustment (or \textsc{GGchem}) can be found on GitHub: \url{https://github.com/DavidDahlbudding/chelio}.\\



\bibliographystyle{mnras}
\bibliography{example} 

@string{june = {June}}

@article{Abel:2011,
 adsurl = {https://ui.adsabs.harvard.edu/abs/2011JPCA..115.6805A},
 author = {{Abel}, Martin and {Frommhold}, Lothar and {Li}, Xiaoping and {Hunt}, Katharine L.~C.},
 doi = {10.1021/jp109441f},
 journal = {Journal of Physical Chemistry A},
 month = {June},
 number = {25},
 pages = {6805-6812},
 title = {{Collision-Induced Absorption by H2Pairs: From Hundreds to Thousands of Kelvin}},
 volume = {115},
 year = {2011}
}

@article{Avila:2021,
 adsurl = {https://ui.adsabs.harvard.edu/abs/2021IJAsB..20..300A},
 archiveprefix = {arXiv},
 author = {{{\'A}vila}, Patricio Javier and {Grassi}, Tommaso and {Bovino}, Stefano and {Chiavassa}, Andrea and {Ercolano}, Barbara and {Danielache}, Sebastian Oscar and {Simoncini}, Eugenio},
 doi = {10.1017/S1473550421000173},
 eprint = {2106.08349},
 journal = {International Journal of Astrobiology},
 month = {August},
 number = {4},
 pages = {300-311},
 primaryclass = {astro-ph.EP},
 title = {{Presence of water on exomoons orbiting free-floating planets: a case study}},
 volume = {20},
 year = {2021}
}

@inproceedings{Baranov:2003,
 address = {Dordrecht},
 author = {Baranov, Y. I. and Fraser, Gerald T. and Lafferty, W. J. and Vigasin, A. A.},
 booktitle = {Weakly Interacting Molecular Pairs: Unconventional Absorbers of Radiation in the Atmosphere},
 doi = {10.1007/978-94-010-0025-3_12},
 editor = {Camy-Peyret, Claude and Vigasin, Andrei A.},
 pages = {149--158},
 publisher = {Springer Netherlands},
 title = {{Collision-Induced Absorption in The CO$_2$ Fermi Triad for Temperatures from 211K to 296 K}},
 year = {2003}
}

@article{Baranov:2004,
 adsurl = {https://ui.adsabs.harvard.edu/abs/2004JMoSp.228..432B},
 author = {{Baranov}, Yuri I. and {Lafferty}, W.~J. and {Fraser}, G.~T.},
 doi = {10.1016/j.jms.2004.04.010},
 journal = {Journal of Molecular Spectroscopy},
 month = {December},
 number = {2},
 pages = {432-440},
 title = {{Infrared spectrum of the continuum and dimer absorption in the vicinity of the O $_{2}$ vibrational fundamental in O $_{2}$/CO $_{2}$ mixtures}},
 volume = {228},
 year = {2004}
}

@article{Baranov:2005,
 adsurl = {https://ui.adsabs.harvard.edu/abs/2005JMoSp.233..160B},
 author = {{Baranov}, Yu. I. and {Lafferty}, W.~J. and {Fraser}, G.~T.},
 doi = {10.1016/j.jms.2005.06.008},
 journal = {Journal of Molecular Spectroscopy},
 month = {September},
 number = {1},
 pages = {160-163},
 title = {{Investigation of collision-induced absorption in the vibrational fundamental bands of O $_{2}$ and N $_{2}$ at elevated temperatures}},
 volume = {233},
 year = {2005}
}

@article{Borysow:2002,
 adsurl = {https://ui.adsabs.harvard.edu/abs/2002A&A...390..779B},
 author = {{Borysow}, A.},
 doi = {10.1051/0004-6361:20020555},
 journal = {\aap},
 month = {August},
 pages = {779-782},
 title = {{Collision-induced absorption coefficients of H$_{2}$ pairs at temperatures from 60 K to 1000 K}},
 volume = {390},
 year = {2002}
}

@article{Borysow:Frommhold:1986a,
 adsurl = {https://ui.adsabs.harvard.edu/abs/1986ApJ...304..849B},
 author = {{Borysow}, A. and {Frommhold}, L.},
 doi = {10.1086/164221},
 journal = {\apj},
 month = {May},
 pages = {849},
 title = {{Theoretical Collision-induced Rototranslational Absorption Spectra for the Outer Planets: H 2-CH 4 Pairs}},
 volume = {304},
 year = {1986}
}

@article{Borysow:Frommhold:1986b,
 adsurl = {https://ui.adsabs.harvard.edu/abs/1986ApJ...303..495B},
 author = {{Borysow}, A. and {Frommhold}, L.},
 doi = {10.1086/164096},
 journal = {\apj},
 month = {April},
 pages = {495},
 title = {{Theoretical Collision-induced Rototranslational Absorption Spectra for Modeling Titan's Atmosphere: H 2--N 2 Pairs}},
 volume = {303},
 year = {1986}
}

@article{Borysow:Frommhold:1987,
 adsurl = {https://ui.adsabs.harvard.edu/abs/1987ApJ...318..940B},
 author = {{Borysow}, Aleksandra and {Frommhold}, Lothar},
 doi = {10.1086/165426},
 journal = {\apj},
 month = {July},
 pages = {940},
 title = {{Collision-induced Rototranslational Absorption Spectra of CH 4-CH 4 Pairs at Temperatures from 50 to 300 K}},
 volume = {318},
 year = {1987}
}

@article{Borysow:Tang:1993,
 adsurl = {https://ui.adsabs.harvard.edu/abs/1993Icar..105..175B},
 author = {{Borysow}, Aleksandra and {Tang}, Chunmei},
 doi = {10.1006/icar.1993.1117},
 journal = {\icarus},
 month = {September},
 number = {1},
 pages = {175-183},
 title = {{Far Infrared CIA Spectra of N $_{2}$-CH $_{4}$ Pairs for Modeling of Titan's Atmosphere}},
 volume = {105},
 year = {1993}
}

@ARTICLE{Brachmann:2025,
       author = {{Brachmann}, Caroline and {Noack}, Lena and {Baumeister}, Philipp Alexander and {Sohl}, Frank},
        title = "{Distinct types of C-H-O-N atmospheres and surface pressures depending on melt redox state and outgassing efficiency}",
      journal = {\icarus},
         year = 2025,
        month = mar,
       volume = {429},
          eid = {116450},
        pages = {116450},
          doi = {10.1016/j.icarus.2024.116450},
       adsurl = {https://ui.adsabs.harvard.edu/abs/2025Icar..42916450B}
}

@inproceedings{Chase:1982,
 adsurl = {https://ui.adsabs.harvard.edu/abs/1982JPhCS..11..695C},
 author = {{Chase}, M.~W. and {Curnutt}, J.~L. and {Downey}, J.~R. and {McDonald}, R.~A. and {Syverud}, A.~N. and {Valenzuela}, E.~A.},
 booktitle = {Journal of Physics Conference Series},
 month = {January},
 pages = {695-940},
 publisher = {IOP},
 series = {Journal of Physics Conference Series},
 title = {{JANAF Thermochemical Tables, 1982 Supplement}},
 volume = {11},
 year = {1982}
}

@book{Chase:1986,
 adsurl = {https://ui.adsabs.harvard.edu/abs/1986jtt..book.....C},
 author = {{Chase}, M.~W.},
 title = {{JANAF thermochemical tables}},
 year = {1986}
}

@article{Chaverot:2025,
 adsurl = {https://ui.adsabs.harvard.edu/abs/2025A&A...702A.137C},
 archiveprefix = {arXiv},
 author = {{Chaverot}, Guillaume and {Turbet}, Martin and {Tran}, Ha and {Hartmann}, Jean-Michel and {Campargue}, Alain and {Mondelain}, Didier and {Bolmont}, Emeline},
 doi = {10.1051/0004-6361/202555762},
 eid = {A137},
 eprint = {2508.18049},
 journal = {\aap},
 month = {October},
 pages = {A137},
 primaryclass = {astro-ph.EP},
 title = {{SpeCT: A state-of-the-art tool to calculate correlated-k tables and continua of CO$_{2}$-H$_{2}$O-N$_{2}$ gas mixtures}},
 volume = {702},
 year = {2025}
}

@article{Chistikov:2019,
 adsurl = {https://ui.adsabs.harvard.edu/abs/2019JChPh.151s4106C},
 author = {{Chistikov}, Daniil N. and {Finenko}, Artem A. and {Lokshtanov}, Sergei E. and {Petrov}, Sergey V. and {Vigasin}, Andrey A.},
 doi = {10.1063/1.5125756},
 eid = {194106},
 journal = {\jcp},
 month = {November},
 number = {19},
 pages = {194106},
 title = {{Simulation of collision-induced absorption spectra based on classical trajectories and ab initio potential and induced dipole surfaces. I. Case study of N$_{2}$-N$_{2}$ rototranslational band}},
 volume = {151},
 year = {2019}
}

@article{Dahlbüdding:2026,
 adsurl = {https://ui.adsabs.harvard.edu/abs/2026MNRAS.548ag243D},
 archiveprefix = {arXiv},
 author = {{Dahlb{\"u}dding}, David and {Grassi}, Tommaso and {Molaverdikhani}, Karan and {Roccetti}, Giulia and {Ercolano}, Barbara and {Braun}, Dieter and {Caselli}, Paola},
 doi = {10.1093/mnras/stag243},
 eid = {stag243},
 eprint = {2602.05378},
 journal = {\mnras},
 month = {May},
 number = {2},
 pages = {stag243},
 primaryclass = {astro-ph.EP},
 title = {{Habitability of tidally heated H$_{2}$-dominated exomoons around free-floating planets}},
 volume = {548},
 year = {2026}
}

@article{deKok:2010,
 adsurl = {https://ui.adsabs.harvard.edu/abs/2010Icar..209..854D},
 author = {{de Kok}, R. and {Irwin}, P.~G.~J. and {Teanby}, N.~A.},
 doi = {10.1016/j.icarus.2010.06.035},
 journal = {\icarus},
 month = {October},
 number = {2},
 pages = {854-857},
 title = {{Far-infrared opacity sources in Titan{\textquoteright}s troposphere reconsidered}},
 volume = {209},
 year = {2010}
}

@book{dePater:Lissauer:2015,
 adsurl = {https://ui.adsabs.harvard.edu/abs/2015plsc.book.....D},
 author = {{de Pater}, Imke and {Lissauer}, Jack J.},
 title = {{Planetary Sciences}},
 year = {2015}
}

@article{Fakhardji:2022,
 adsurl = {https://ui.adsabs.harvard.edu/abs/2022JQSRT.28308148F},
 author = {{Fakhardji}, Wissam and {Boulet}, Christian and {Tran}, Ha and {Hartmann}, Jean-Michel},
 doi = {10.1016/j.jqsrt.2022.108148},
 eid = {108148},
 journal = {\jqsrt},
 month = {June},
 pages = {108148},
 title = {{Direct calculations of the CH$_{4}$+CO$_{2}$ far infrared collision-induced absorption}},
 volume = {283},
 year = {2022}
}

@article{Finenko:2022,
 adsurl = {https://ui.adsabs.harvard.edu/abs/2022ApJS..258...33F},
 author = {{Finenko}, Artem A. and {B{\'e}zard}, Bruno and {Gordon}, Iouli E. and {Chistikov}, Daniil N. and {Lokshtanov}, Sergei E. and {Petrov}, Sergey V. and {Vigasin}, Andrey A.},
 doi = {10.3847/1538-4365/ac36d3},
 eid = {33},
 journal = {\apjs},
 month = {February},
 number = {2},
 pages = {33},
 title = {{Trajectory-based Simulation of Far-infrared Collision-induced Absorption Profiles of CH$_{4}$-N$_{2}$ for Modeling Titan's Atmosphere}},
 volume = {258},
 year = {2022}
}

@article{Fletcher:2018,
 adsurl = {https://ui.adsabs.harvard.edu/abs/2018ApJS..235...24F},
 archiveprefix = {arXiv},
 author = {{Fletcher}, Leigh N. and {Gustafsson}, Magnus and {Orton}, Glenn S.},
 doi = {10.3847/1538-4365/aaa07a},
 eid = {24},
 eprint = {1712.02813},
 journal = {\apjs},
 month = {March},
 number = {1},
 pages = {24},
 primaryclass = {astro-ph.EP},
 title = {{Hydrogen Dimers in Giant-planet Infrared Spectra}},
 volume = {235},
 year = {2018}
}

@book{Frommhold:2006,
 author = {Frommhold, Lothar},
 collection = {Cambridge Monographs on Atomic, Molecular and Chemical Physics},
 note = {originally published 1994},
 place = {Cambridge},
 publisher = {Cambridge University Press},
 series = {Cambridge Monographs on Atomic, Molecular and Chemical Physics},
 title = {Collision-induced Absorption in Gases},
 year = {2006}
}

@article{Gharib-Nezhad:2024,
 adsurl = {https://ui.adsabs.harvard.edu/abs/2024RASTI...3...44G},
 archiveprefix = {arXiv},
 author = {{Gharib-Nezhad}, Ehsan (Sam) and {Batalha}, Natasha E. and {Chubb}, Katy and {Freedman}, Richard and {Gordon}, Iouli E. and {Gamache}, Robert R. and {Hargreaves}, Robert J. and {Lewis}, Nikole K. and {Tennyson}, Jonathan and {Yurchenko}, Sergei N.},
 doi = {10.1093/rasti/rzad058},
 eprint = {2401.03056},
 journal = {RAS Techniques and Instruments},
 month = {January},
 number = {1},
 pages = {44-55},
 primaryclass = {astro-ph.EP},
 title = {{The impact of spectral line wing cut-off: recommended standard method with application to MAESTRO opacity data base}},
 volume = {3},
 year = {2024}
}

@ARTICLE{Goldblatt:2013,
       author = {{Goldblatt}, Colin and {Robinson}, Tyler D. and {Zahnle}, Kevin J. and {Crisp}, David},
        title = "{Low simulated radiation limit for runaway greenhouse climates}",
      journal = {Nature Geoscience},
         year = 2013,
        month = aug,
       volume = {6},
       number = {8},
        pages = {661-667},
          doi = {10.1038/ngeo1892},
       adsurl = {https://ui.adsabs.harvard.edu/abs/2013NatGe...6..661G}
}

@article{Gordon:2021,
 adsurl = {https://ui.adsabs.harvard.edu/abs/2022JQSRT.27707949G},
 author = {{Gordon}, I.~E. and {Rothman}, L.~S. and {Hargreaves}, R.~J. and {Hashemi}, R. and {Karlovets}, E.~V. and {Skinner}, F.~M. and {Conway}, E.~K. and {Hill}, C. and {Kochanov}, R.~V. and {Tan}, Y. and {Wcis{\l}o}, P. and {Finenko}, A.~A. and {Nelson}, K. and {Bernath}, P.~F. and {Birk}, M. and {Boudon}, V. and {Campargue}, A. and {Chance}, K.~V. and {Coustenis}, A. and {Drouin}, B.~J. and {Flaud}, J.-M. and {Gamache}, R.~R. and {Hodges}, J.~T. and {Jacquemart}, D. and {Mlawer}, E.~J. and {Nikitin}, A.~V. and {Perevalov}, V.~I. and {Rotger}, M. and {Tennyson}, J. and {Toon}, G.~C. and {Tran}, H. and {Tyuterev}, V.~G. and {Adkins}, E.~M. and {Baker}, A. and {Barbe}, A. and {Can{\`e}}, E. and {Cs{\'a}sz{\'a}r}, A.~G. and {Dudaryonok}, A. and {Egorov}, O. and {Fleisher}, A.~J. and {Fleurbaey}, H. and {Foltynowicz}, A. and {Furtenbacher}, T. and {Harrison}, J.~J. and {Hartmann}, J.-M. and {Horneman}, V.-M. and {Huang}, X. and {Karman}, T. and {Karns}, J. and {Kassi}, S. and {Kleiner}, I. and {Kofman}, V. and {Kwabia-Tchana}, F. and {Lavrentieva}, N.~N. and {Lee}, T.~J. and {Long}, D.~A. and {Lukashevskaya}, A.~A. and {Lyulin}, O.~M. and {Makhnev}, V. Yu. and {Matt}, W. and {Massie}, S.~T. and {Melosso}, M. and {Mikhailenko}, S.~N. and {Mondelain}, D. and {M{\"u}ller}, H.~S.~P. and {Naumenko}, O.~V. and {Perrin}, A. and {Polyansky}, O.~L. and {Raddaoui}, E. and {Raston}, P.~L. and {Reed}, Z.~D. and {Rey}, M. and {Richard}, C. and {T{\'o}bi{\'a}s}, R. and {Sadiek}, I. and {Schwenke}, D.~W. and {Starikova}, E. and {Sung}, K. and {Tamassia}, F. and {Tashkun}, S.~A. and {Vander Auwera}, J. and {Vasilenko}, I.~A. and {Vigasin}, A.~A. and {Villanueva}, G.~L. and {Vispoel}, B. and {Wagner}, G. and {Yachmenev}, A. and {Yurchenko}, S.~N.},
 doi = {10.1016/j.jqsrt.2021.107949},
 eid = {107949},
 journal = {\jqsrt},
 month = {January},
 pages = {107949},
 title = {{The HITRAN2020 molecular spectroscopic database}},
 volume = {277},
 year = {2022}
}

@article{Grimm:2021,
 adsurl = {https://ui.adsabs.harvard.edu/abs/2021ApJS..253...30G},
 archiveprefix = {arXiv},
 author = {{Grimm}, Simon L. and {Malik}, Matej and {Kitzmann}, Daniel and {Guzm{\'a}n-Mesa}, Andrea and {Hoeijmakers}, H. Jens and {Fisher}, Chloe and {Mendon{\c{c}}a}, Jo{\~a}o M. and {Yurchenko}, Sergey N. and {Tennyson}, Jonathan and {Alesina}, Fabien and {Buchschacher}, Nicolas and {Burnier}, Julien and {Segransan}, Damien and {Kurucz}, Robert L. and {Heng}, Kevin},
 doi = {10.3847/1538-4365/abd773},
 eid = {30},
 eprint = {2101.02005},
 journal = {\apjs},
 month = {March},
 number = {1},
 pages = {30},
 primaryclass = {astro-ph.EP},
 title = {{HELIOS-K 2.0 Opacity Calculator and Open-source Opacity Database for Exoplanetary Atmospheres}},
 volume = {253},
 year = {2021}
}

@article{Gruszka:Borysow:1997,
 adsurl = {https://ui.adsabs.harvard.edu/abs/1997Icar..129..172G},
 author = {{Gruszka}, Marcin and {Borysow}, Aleksandra},
 doi = {10.1006/icar.1997.5773},
 journal = {\icarus},
 month = {September},
 number = {1},
 pages = {172-177},
 title = {{Roto-Translational Collision-Induced Absorption of CO $_{2}$for the Atmosphere of Venus at Frequencies from 0 to 250 cm $^{-1}$, at Temperatures from 200 to 800 K}},
 volume = {129},
 year = {1997}
}

@article{Guillot:2005,
 adsurl = {https://ui.adsabs.harvard.edu/abs/2005AREPS..33..493G},
 archiveprefix = {arXiv},
 author = {{Guillot}, Tristan},
 doi = {10.1146/annurev.earth.32.101802.120325},
 eprint = {astro-ph/0502068},
 journal = {Annual Review of Earth and Planetary Sciences},
 month = {January},
 pages = {493-530},
 primaryclass = {astro-ph},
 title = {{THE INTERIORS OF GIANT PLANETS: Models and Outstanding Questions}},
 volume = {33},
 year = {2005}
}

@article{Hartmann:2017,
 adsurl = {https://ui.adsabs.harvard.edu/abs/2017JGRD..122.2419H},
 author = {{Hartmann}, J.-M. and {Boulet}, C. and {Toon}, G.~C.},
 doi = {10.1002/2016JD025677},
 journal = {Journal of Geophysical Research (Atmospheres)},
 month = {February},
 number = {4},
 pages = {2419-2428},
 title = {{Collision-induced absorption by N$_{2}$ near 2.16 {\textmu}m: Calculations, model, and consequences for atmospheric remote sensing}},
 volume = {122},
 year = {2017}
}

@article{Hedges:2016,
 adsurl = {https://ui.adsabs.harvard.edu/abs/2016MNRAS.458.1427H},
 archiveprefix = {arXiv},
 author = {{Hedges}, Christina and {Madhusudhan}, Nikku},
 doi = {10.1093/mnras/stw278},
 eprint = {1602.00751},
 journal = {\mnras},
 month = {May},
 number = {2},
 pages = {1427-1449},
 primaryclass = {astro-ph.EP},
 title = {{Effect of pressure broadening on molecular absorption cross sections in exoplanetary atmospheres}},
 volume = {458},
 year = {2016}
}

@article{Hu:2015,
 adsurl = {https://ui.adsabs.harvard.edu/abs/2015ApJ...807....8H},
 archiveprefix = {arXiv},
 author = {{Hu}, Renyu and {Seager}, Sara and {Yung}, Yuk L.},
 doi = {10.1088/0004-637X/807/1/8},
 eid = {8},
 eprint = {1505.02221},
 journal = {\apj},
 month = {July},
 number = {1},
 pages = {8},
 primaryclass = {astro-ph.EP},
 title = {{Helium Atmospheres on Warm Neptune- and Sub-Neptune-sized Exoplanets and Applications to GJ 436b}},
 volume = {807},
 year = {2015}
}

@article{Johnson:2024,
 adsurl = {https://ui.adsabs.harvard.edu/abs/2024JQSRT.32909189J},
 author = {{Johnson}, R. and {Bernath}, P. and {Billinghurst}, B. and {Zhao}, J.},
 doi = {10.1016/j.jqsrt.2024.109189},
 eid = {109189},
 journal = {\jqsrt},
 month = {December},
 pages = {109189},
 title = {{Collision induced absorption spectra of N$_{2}$ and CH$_{4}$}},
 volume = {329},
 year = {2024}
}

@article{Karman:2019,
 adsurl = {https://ui.adsabs.harvard.edu/abs/2019Icar..328..160K},
 author = {{Karman}, Tijs and {Gordon}, Iouli E. and {van der Avoird}, Ad and {Baranov}, Yury I. and {Boulet}, Christian and {Drouin}, Brian J. and {Groenenboom}, Gerrit C. and {Gustafsson}, Magnus and {Hartmann}, Jean-Michel and {Kurucz}, Robert L. and {Rothman}, Laurence S. and {Sun}, Kang and {Sung}, Keeyoon and {Thalman}, Ryan and {Tran}, Ha and {Wishnow}, Edward H. and {Wordsworth}, Robin and {Vigasin}, Andrey A. and {Volkamer}, Rainer and {van der Zande}, Wim J.},
 doi = {10.1016/j.icarus.2019.02.034},
 journal = {\icarus},
 month = {August},
 pages = {160-175},
 title = {{Update of the HITRAN collision-induced absorption section}},
 volume = {328},
 year = {2019}
}

@ARTICLE{Kasting:1988,
       author = {{Kasting}, J.~F.},
        title = "{Runaway and moist greenhouse atmospheres and the evolution of Earth and Venus}",
      journal = {\icarus},
         year = 1988,
        month = jun,
       volume = {74},
       number = {3},
        pages = {472-494},
          doi = {10.1016/0019-1035(88)90116-9},
       adsurl = {https://ui.adsabs.harvard.edu/abs/1988Icar...74..472K}
}

@article{Kasting:2013,
 adsurl = {https://ui.adsabs.harvard.edu/abs/2013Sci...339...44K},
 author = {{Kasting}, James F.},
 doi = {10.1126/science.1232662},
 journal = {Science},
 month = {January},
 number = {6115},
 pages = {44},
 title = {{How Was Early Earth Kept Warm?}},
 volume = {339},
 year = {2013}
}

@dataset{kitzmann:2025,
  author       = {Kitzmann, Daniel and Malik, Matej},
  title        = {Basic input data for the HELIOS atmosphere model},
  month        = oct,
  year         = 2025,
  publisher    = {Zenodo},
  doi          = {10.5281/zenodo.17425932},
  url          = {https://doi.org/10.5281/zenodo.17425932},
}

@article{Lafferty:1996,
 adsurl = {https://ui.adsabs.harvard.edu/abs/1996ApOpt..35.5911L},
 author = {{Lafferty}, Walter J. and {Solodov}, Alexander M. and {Weber}, Alfons and {Olson}, Wm. Bruce and {Hartmann}, Jean-Michel},
 doi = {10.1364/AO.35.005911},
 journal = {\ao},
 month = {October},
 number = {30},
 pages = {5911-5917},
 title = {{Infrared collision-induced absorption by N 2 near 4.3 m for atmospheric applications: measurements and empirical modeling}},
 volume = {35},
 year = {1996}
}

@article{Lammer:2025,
 adsurl = {https://ui.adsabs.harvard.edu/abs/2025NatAs...9.1022L},
 author = {{Lammer}, Helmut and {Scherf}, Manuel and {Erkaev}, Nikolai V. and {Kubyshkina}, Daria and {Gorbunova}, Kseniia D. and {Fossati}, Luca and {Woitke}, Peter},
 doi = {10.1038/s41550-025-02550-6},
 journal = {Nature Astronomy},
 month = {July},
 pages = {1022-1030},
 title = {{Earth-mass planets with He atmospheres in the habitable zone of Sun-like stars}},
 volume = {9},
 year = {2025}
}

@ARTICLE{Ledoux:1947,
       author = {{Ledoux}, P.},
        title = "{Stellar Models with Convection and with Discontinuity of the Mean Molecular Weight}",
      journal = {\apj},
         year = 1947,
        month = mar,
       volume = {105},
        pages = {305},
          doi = {10.1086/144905},
       adsurl = {https://ui.adsabs.harvard.edu/abs/1947ApJ...105..305L}
}

@article{Malik:2017,
 adsurl = {https://ui.adsabs.harvard.edu/abs/2017AJ....153...56M},
 archiveprefix = {arXiv},
 author = {{Malik}, Matej and {Grosheintz}, Luc and {Mendon{\c{c}}a}, Jo{\~a}o M. and {Grimm}, Simon L. and {Lavie}, Baptiste and {Kitzmann}, Daniel and {Tsai}, Shang-Min and {Burrows}, Adam and {Kreidberg}, Laura and {Bedell}, Megan and {Bean}, Jacob L. and {Stevenson}, Kevin B. and {Heng}, Kevin},
 doi = {10.3847/1538-3881/153/2/56},
 eid = {56},
 eprint = {1606.05474},
 journal = {\aj},
 month = {February},
 number = {2},
 pages = {56},
 primaryclass = {astro-ph.EP},
 title = {{HELIOS: An Open-source, GPU-accelerated Radiative Transfer Code for Self-consistent Exoplanetary Atmospheres}},
 volume = {153},
 year = {2017}
}

@article{Malik:2019a,
 adsurl = {https://ui.adsabs.harvard.edu/abs/2019AJ....157..170M},
 archiveprefix = {arXiv},
 author = {{Malik}, Matej and {Kitzmann}, Daniel and {Mendon{\c{c}}a}, Jo{\~a}o M. and {Grimm}, Simon L. and {Marleau}, Gabriel-Dominique and {Linder}, Esther F. and {Tsai}, Shang-Min and {Heng}, Kevin},
 doi = {10.3847/1538-3881/ab1084},
 eid = {170},
 eprint = {1903.06794},
 journal = {\aj},
 month = {May},
 number = {5},
 pages = {170},
 primaryclass = {astro-ph.EP},
 title = {{Self-luminous and Irradiated Exoplanetary Atmospheres Explored with HELIOS}},
 volume = {157},
 year = {2019}
}

@article{Malik:2019b,
 adsurl = {https://ui.adsabs.harvard.edu/abs/2019ApJ...886..142M},
 archiveprefix = {arXiv},
 author = {{Malik}, Matej and {Kempton}, Eliza M.-R. and {Koll}, Daniel D.~B. and {Mansfield}, Megan and {Bean}, Jacob L. and {Kite}, Edwin},
 doi = {10.3847/1538-4357/ab4a05},
 eid = {142},
 eprint = {1907.13135},
 journal = {\apj},
 month = {December},
 number = {2},
 pages = {142},
 primaryclass = {astro-ph.EP},
 title = {{Analyzing Atmospheric Temperature Profiles and Spectra of M Dwarf Rocky Planets}},
 volume = {886},
 year = {2019}
}

@article{Mlawer:2023,
 adsurl = {https://ui.adsabs.harvard.edu/abs/2023JQSRT.30608645M},
 author = {{Mlawer}, E.~J. and {Cady-Pereira}, K.~E. and {Mascio}, J. and {Gordon}, I.~E.},
 doi = {10.1016/j.jqsrt.2023.108645},
 eid = {108645},
 journal = {\jqsrt},
 month = {September},
 pages = {108645},
 title = {{The inclusion of the MT\_CKD water vapor continuum model in the HITRAN molecular spectroscopic database}},
 volume = {306},
 year = {2023}
}

@article{Molliere:2015,
 adsurl = {https://ui.adsabs.harvard.edu/abs/2015ApJ...813...47M},
 archiveprefix = {arXiv},
 author = {{Molli{\`e}re}, P. and {van Boekel}, R. and {Dullemond}, C. and {Henning}, Th. and {Mordasini}, C.},
 doi = {10.1088/0004-637X/813/1/47},
 eid = {47},
 eprint = {1509.07523},
 journal = {\apj},
 month = {November},
 number = {1},
 pages = {47},
 primaryclass = {astro-ph.EP},
 title = {{Model Atmospheres of Irradiated Exoplanets: The Influence of Stellar Parameters, Metallicity, and the C/O Ratio}},
 volume = {813},
 year = {2015}
}

@article{MolLous:2022,
 adsurl = {https://ui.adsabs.harvard.edu/abs/2022NatAs...6..819M},
 archiveprefix = {arXiv},
 author = {{Mol Lous}, Marit and {Helled}, Ravit and {Mordasini}, Christoph},
 doi = {10.1038/s41550-022-01699-8},
 eprint = {2206.13859},
 journal = {Nature Astronomy},
 month = {June},
 pages = {819-827},
 primaryclass = {astro-ph.EP},
 title = {{Potential long-term habitable conditions on planets with primordial H-He atmospheres}},
 volume = {6},
 year = {2022}
}

@ARTICLE{Nakajima:1992,
       author = {{Nakajima}, Shinichi and {Hayashi}, Yoshi-Yuki and {Abe}, Yutaka},
        title = "{A study on the 'runaway greenhouse effect' with a one-dimensional radiative-convective equilibrium model}",
      journal = {Journal of the Atmospheric Sciences},
         year = 1992,
        month = dec,
       volume = {49},
       number = {23},
        pages = {2256-2266},
          doi = {10.1175/1520-0469(1992)049<2256:ASOTGE>2.0.CO;2},
       adsurl = {https://ui.adsabs.harvard.edu/abs/1992JAtS...49.2256N}
}

@article{Nicholls:2024,
 adsurl = {https://ui.adsabs.harvard.edu/abs/2024JGRE..12908576N},
 archiveprefix = {arXiv},
 author = {{Nicholls}, Harrison and {Lichtenberg}, Tim and {Bower}, Dan J. and {Pierrehumbert}, Raymond},
 doi = {10.1029/2024JE008576},
 eprint = {2411.19137},
 journal = {Journal of Geophysical Research (Planets)},
 month = {December},
 number = {12},
 pages = {2024JE008576},
 primaryclass = {astro-ph.EP},
 title = {{Magma Ocean Evolution at Arbitrary Redox State}},
 volume = {129},
 year = {2024}
}

@article{Nicholls:2025a,
 adsurl = {https://ui.adsabs.harvard.edu/abs/2025MNRAS.536.2957N},
 archiveprefix = {arXiv},
 author = {{Nicholls}, Harrison and {Pierrehumbert}, Raymond T. and {Lichtenberg}, Tim and {Soucasse}, Laurent and {Smeets}, Stef},
 doi = {10.1093/mnras/stae2772},
 eprint = {2412.11987},
 journal = {\mnras},
 month = {January},
 number = {3},
 pages = {2957-2971},
 primaryclass = {astro-ph.EP},
 title = {{Convective shutdown in the atmospheres of lava worlds}},
 volume = {536},
 year = {2025}
}

@article{Nicholls:2025b,
 adsurl = {https://ui.adsabs.harvard.edu/abs/2025JOSS...10.7726N},
 archiveprefix = {arXiv},
 author = {{Nicholls}, Harrison and {Pierrehumbert}, Raymond and {Lichtenberg}, Tim},
 doi = {10.21105/joss.07726},
 eid = {7726},
 eprint = {2506.00091},
 journal = {The Journal of Open Source Software},
 month = {May},
 number = {109},
 pages = {7726},
 primaryclass = {astro-ph.IM},
 title = {{AGNI: A radiative-convective model for lava planet atmospheres}},
 volume = {10},
 year = {2025}
}

@ARTICLE{Ozak:2016,
       author = {{Ozak}, N. and {Aharonson}, O. and {Halevy}, I.},
        title = "{Radiative transfer in CO$_{2}$-rich atmospheres: 1. Collisional line mixing implies a colder early Mars}",
      journal = {Journal of Geophysical Research (Planets)},
         year = 2016,
        month = jun,
       volume = {121},
       number = {6},
        pages = {965-985},
          doi = {10.1002/2015JE004871},
       adsurl = {https://ui.adsabs.harvard.edu/abs/2016JGRE..121..965O}
}

@article{Polyansky:2018,
 adsurl = {https://ui.adsabs.harvard.edu/abs/2018MNRAS.480.2597P},
 archiveprefix = {arXiv},
 author = {{Polyansky}, Oleg L. and {Kyuberis}, Aleksandra A. and {Zobov}, Nikolai F. and {Tennyson}, Jonathan and {Yurchenko}, Sergei N. and {Lodi}, Lorenzo},
 doi = {10.1093/mnras/sty1877},
 eprint = {1807.04529},
 journal = {\mnras},
 month = {October},
 number = {2},
 pages = {2597-2608},
 primaryclass = {astro-ph.EP},
 title = {{ExoMol molecular line lists XXX: a complete high-accuracy line list for water}},
 volume = {480},
 year = {2018}
}

@article{Ramirez:2014,
 adsurl = {https://ui.adsabs.harvard.edu/abs/2014NatGe...7...59R},
 archiveprefix = {arXiv},
 author = {{Ramirez}, Ramses M. and {Kopparapu}, Ravi and {Zugger}, Michael E. and {Robinson}, Tyler D. and {Freedman}, Richard and {Kasting}, James F.},
 doi = {10.1038/ngeo2000},
 eprint = {1405.6701},
 journal = {Nature Geoscience},
 month = {January},
 number = {1},
 pages = {59-63},
 primaryclass = {astro-ph.EP},
 title = {{Warming early Mars with CO$_{2}$ and H$_{2}$}},
 volume = {7},
 year = {2014}
}

@article{Richard:2011,
 adsurl = {https://ui.adsabs.harvard.edu/abs/2012JQSRT.113.1276R},
 author = {{Richard}, C. and {Gordon}, I.~E. and {Rothman}, L.~S. and {Abel}, M. and {Frommhold}, L. and {Gustafsson}, M. and {Hartmann}, J.-M. and {Hermans}, C. and {Lafferty}, W.~J. and {Orton}, G.~S. and {Smith}, K.~M. and {Tran}, H.},
 doi = {10.1016/j.jqsrt.2011.11.004},
 journal = {\jqsrt},
 month = {July},
 number = {11},
 pages = {1276-1285},
 title = {{New section of the HITRAN database: Collision-induced absorption (CIA)}},
 volume = {113},
 year = {2012}
}

@article{Roccetti:2023,
 adsurl = {https://ui.adsabs.harvard.edu/abs/2023IJAsB..22..317R},
 archiveprefix = {arXiv},
 author = {{Roccetti}, Giulia and {Grassi}, Tommaso and {Ercolano}, Barbara and {Molaverdikhani}, Karan and {Crida}, Aur{\'e}lien and {Braun}, Dieter and {Chiavassa}, Andrea},
 doi = {10.1017/S1473550423000046},
 eprint = {2302.04946},
 journal = {International Journal of Astrobiology},
 month = {August},
 number = {4},
 pages = {317-346},
 primaryclass = {astro-ph.EP},
 title = {{Presence of liquid water during the evolution of exomoons orbiting ejected free-floating planets}},
 volume = {22},
 year = {2023}
}

@article{Seager:2010,
 adsurl = {https://ui.adsabs.harvard.edu/abs/2010ARA&A..48..631S},
 archiveprefix = {arXiv},
 author = {{Seager}, Sara and {Deming}, Drake},
 doi = {10.1146/annurev-astro-081309-130837},
 eprint = {1005.4037},
 journal = {\araa},
 month = {September},
 pages = {631-672},
 primaryclass = {astro-ph.EP},
 title = {{Exoplanet Atmospheres}},
 volume = {48},
 year = {2010}
}

@article{Stevenson:1999,
 adsurl = {https://ui.adsabs.harvard.edu/abs/1999Natur.400...32S},
 author = {{Stevenson}, David J.},
 doi = {10.1038/21811},
 journal = {\nat},
 month = {July},
 number = {6739},
 pages = {32},
 title = {{Life-sustaining planets in interstellar space?}},
 volume = {400},
 year = {1999}
}

@inproceedings{Sung:2016,
 adsurl = {https://ui.adsabs.harvard.edu/abs/2016DPS....4842411S},
 author = {{Sung}, Keeyoon and {Wishnow}, Ed and {Venkataraman}, Malathy and {Brown}, Linda R. and {Ozier}, Irving and {Benner}, D. Chris and {Crawford}, Tomithy J. and {Mantz}, Arlan and {Smith}, Mary-Ann H.},
 booktitle = {AAS/Division for Planetary Sciences Meeting Abstracts \#48},
 eid = {424.11},
 month = {October},
 pages = {424.11},
 series = {AAS/Division for Planetary Sciences Meeting Abstracts},
 title = {{Progress in the measurement of temperature-dependent N$_{2}$-N$_{2}$ collision-induced absorption and H$_{2}$-broadening of cold and hot CH$_{4}$}},
 volume = {48},
 year = {2016}
}

@article{Terragni:2025,
 adsurl = {https://ui.adsabs.harvard.edu/abs/2025JQSRT.34709631T},
 author = {{Terragni}, J. and {Gordon}, I.~E. and {Adkins}, E.~M. and {Boulet}, C. and {Campargue}, A. and {Chistikov}, D. and {Finenko}, A. and {Finkenzeller}, H. and {Fleurbaey}, H. and {Hargreaves}, R.~J. and {Hanson}, R.~K. and {Hartmann}, J.-M. and {Klingberg}, A. and {Kohler}, E. and {Koroleva}, A.~O. and {Mondelain}, D. and {Piccioni}, G. and {Stefani}, S. and {Strand}, C.~L. and {Tran}, H. and {Turbet}, M. and {Vigasin}, A. and {Vitali}, F. and {Volkamer}, R. and {Wei}, C.},
 doi = {10.1016/j.jqsrt.2025.109631},
 eid = {109631},
 journal = {\jqsrt},
 month = {December},
 pages = {109631},
 title = {{Collision induced absorption in HITRAN2024: Enhanced and improved data for atmospheric and planetary studies}},
 volume = {347},
 year = {2025}
}

@article{Tran:2024,
 adsurl = {https://ui.adsabs.harvard.edu/abs/2024Icar..42216265T},
 author = {{Tran}, H. and {Hartmann}, J.~M. and {Rambinison}, E. and {Turbet}, M.},
 doi = {10.1016/j.icarus.2024.116265},
 eid = {116265},
 journal = {\icarus},
 month = {November},
 pages = {116265},
 title = {{Collision-induced absorptions by pure CO$_{2}$ in the infrared: New measurements in the 1150─4500 cm$^{{\ensuremath{-}}1}$ spectral range and empirical modeling for applications}},
 volume = {422},
 year = {2024}
}

@ARTICLE{Turbet:2017,
       author = {{Turbet}, M. and {Tran}, H.},
        title = "{Comment on ``Radiative Transfer in CO$_{2}$-Rich Atmospheres: 1. Collisional Line Mixing Implies a Colder Early Mars''}",
      journal = {Journal of Geophysical Research (Planets)},
         year = 2017,
        month = nov,
       volume = {122},
       number = {11},
        pages = {2362-2365},
          doi = {10.1002/2017JE005373},
archivePrefix = {arXiv},
       eprint = {1711.03376},
 primaryClass = {astro-ph.EP},
       adsurl = {https://ui.adsabs.harvard.edu/abs/2017JGRE..122.2362T}
}

@article{Turbet:2020,
 adsurl = {https://ui.adsabs.harvard.edu/abs/2020Icar..34613762T},
 archiveprefix = {arXiv},
 author = {{Turbet}, Martin and {Boulet}, Christian and {Karman}, Tijs},
 doi = {10.1016/j.icarus.2020.113762},
 eid = {113762},
 eprint = {1912.05630},
 journal = {\icarus},
 month = {August},
 pages = {113762},
 primaryclass = {astro-ph.EP},
 title = {{Measurements and semi-empirical calculations of CO$_{2}$ + CH$_{4}$ and CO$_{2}$ + H$_{2}$ collision-induced absorption across a wide range of wavelengths and temperatures. Application for the prediction of early Mars surface temperature}},
 volume = {346},
 year = {2020}
}

@article{Vigasin:2014,
 adsurl = {https://ui.adsabs.harvard.edu/abs/2014JQSRT.148...58V},
 author = {{Vigasin}, A.~A.},
 doi = {10.1016/j.jqsrt.2014.06.019},
 journal = {\jqsrt},
 month = {November},
 pages = {58-64},
 title = {{Water vapor continuum: Whether collision-induced absorption is involved?}},
 volume = {148},
 year = {2014}
}

@article{Wei:2024,
 adsurl = {https://ui.adsabs.harvard.edu/abs/2024IJHE...93..364W},
 author = {{Wei}, Chuyu and {Klingberg}, Andrew and {Strand}, Christopher L. and {Hanson}, Ronald K.},
 doi = {10.1016/j.ijhydene.2024.10.318},
 journal = {International Journal of Hydrogen Energy},
 month = {December},
 pages = {364-373},
 title = {{Measurement of hydrogen and nitrogen via collision-induced infrared absorption}},
 volume = {93},
 year = {2024}
}

@article{Whittaker:2022,
 adsurl = {https://ui.adsabs.harvard.edu/abs/2022AJ....164..258W},
 archiveprefix = {arXiv},
 author = {{Whittaker}, Emily A. and {Malik}, Matej and {Ih}, Jegug and {Kempton}, Eliza M.-R. and {Mansfield}, Megan and {Bean}, Jacob L. and {Kite}, Edwin S. and {Koll}, Daniel D.~B. and {Cronin}, Timothy W. and {Hu}, Renyu},
 doi = {10.3847/1538-3881/ac9ab3},
 eid = {258},
 eprint = {2207.08889},
 journal = {\aj},
 month = {December},
 number = {6},
 pages = {258},
 primaryclass = {astro-ph.EP},
 title = {{The Detectability of Rocky Planet Surface and Atmosphere Composition with the JWST: The Case of LHS 3844b}},
 volume = {164},
 year = {2022}
}

@article{Woitke:2018,
 adsurl = {https://ui.adsabs.harvard.edu/abs/2018A&A...614A...1W},
 archiveprefix = {arXiv},
 author = {{Woitke}, P. and {Helling}, Ch. and {Hunter}, G.~H. and {Millard}, J.~D. and {Turner}, G.~E. and {Worters}, M. and {Blecic}, J. and {Stock}, J.~W.},
 doi = {10.1051/0004-6361/201732193},
 eid = {A1},
 eprint = {1712.01010},
 journal = {\aap},
 month = {June},
 pages = {A1},
 primaryclass = {astro-ph.EP},
 title = {{Equilibrium chemistry down to 100 K. Impact of silicates and phyllosilicates on the carbon to oxygen ratio}},
 volume = {614},
 year = {2018}
}

@article{Woitke:2021,
       author = {{Woitke}, P. and {Herbort}, O. and {Helling}, Ch. and {St{\"u}eken}, E. and {Dominik}, M. and {Barth}, P. and {Samra}, D.},
        title = "{Coexistence of CH$_{4}$, CO$_{2}$, and H$_{2}$O in exoplanet atmospheres}",
      journal = {\aap},
         year = 2021,
        month = feb,
       volume = {646},
          eid = {A43},
        pages = {A43},
          doi = {10.1051/0004-6361/202038870},
archivePrefix = {arXiv},
       eprint = {2010.12241},
 primaryClass = {astro-ph.EP},
       adsurl = {https://ui.adsabs.harvard.edu/abs/2021A&A...646A..43W}
}

@article{Wordsworth:2013,
 adsurl = {https://ui.adsabs.harvard.edu/abs/2013Sci...339...64W},
 author = {{Wordsworth}, Robin and {Pierrehumbert}, Raymond},
 doi = {10.1126/science.1225759},
 journal = {Science},
 month = {January},
 number = {6115},
 pages = {64},
 title = {{Hydrogen-Nitrogen Greenhouse Warming in Earth's Early Atmosphere}},
 volume = {339},
 year = {2013}
}

@article{Wordsworth:2017,
 adsurl = {https://ui.adsabs.harvard.edu/abs/2017GeoRL..44..665W},
 archiveprefix = {arXiv},
 author = {{Wordsworth}, R. and {Kalugina}, Y. and {Lokshtanov}, S. and {Vigasin}, A. and {Ehlmann}, B. and {Head}, J. and {Sanders}, C. and {Wang}, H.},
 doi = {10.1002/2016GL071766},
 eprint = {1610.09697},
 journal = {\grl},
 month = {January},
 number = {2},
 pages = {665-671},
 primaryclass = {astro-ph.EP},
 title = {{Transient reducing greenhouse warming on early Mars}},
 volume = {44},
 year = {2017}
}

@article{Yurchenko:2014,
 adsurl = {https://ui.adsabs.harvard.edu/abs/2014MNRAS.440.1649Y},
 archiveprefix = {arXiv},
 author = {{Yurchenko}, Sergei N. and {Tennyson}, Jonathan},
 doi = {10.1093/mnras/stu326},
 eprint = {1401.4852},
 journal = {\mnras},
 month = {May},
 number = {2},
 pages = {1649-1661},
 primaryclass = {astro-ph.EP},
 title = {{ExoMol line lists - IV. The rotation-vibration spectrum of methane up to 1500 K}},
 volume = {440},
 year = {2014}
}

@article{Yurchenko:2017,
 adsurl = {https://ui.adsabs.harvard.edu/abs/2017A&A...605A..95Y},
 archiveprefix = {arXiv},
 author = {{Yurchenko}, Sergei N. and {Amundsen}, David S. and {Tennyson}, Jonathan and {Waldmann}, Ingo P.},
 doi = {10.1051/0004-6361/201731026},
 eid = {A95},
 eprint = {1706.05724},
 journal = {\aap},
 month = {September},
 pages = {A95},
 primaryclass = {astro-ph.EP},
 title = {{A hybrid line list for CH$_{4}$ and hot methane continuum}},
 volume = {605},
 year = {2017}
}

@article{Yurchenko:2020,
 adsurl = {https://ui.adsabs.harvard.edu/abs/2020MNRAS.496.5282Y},
 archiveprefix = {arXiv},
 author = {{Yurchenko}, S.~N. and {Mellor}, Thomas M. and {Freedman}, Richard S. and {Tennyson}, J.},
 doi = {10.1093/mnras/staa1874},
 eprint = {2007.02122},
 journal = {\mnras},
 month = {August},
 number = {4},
 pages = {5282-5291},
 primaryclass = {astro-ph.EP},
 title = {{ExoMol line lists - XXXIX. Ro-vibrational molecular line list for CO$_{2}$}},
 volume = {496},
 year = {2020}
}

@inproceedings{Zahnle:1993,
 adsurl = {https://ui.adsabs.harvard.edu/abs/1993prpl.conf.1305Z},
 author = {{Zahnle}, Kevin},
 booktitle = {Protostars and Planets III},
 editor = {{Levy}, Eugene H. and {Lunine}, Jonathan I.},
 month = {January},
 pages = {1305},
 title = {{Planetary Noble Gases}},
 year = {1993}
}




\appendix

\section{Molecular Opacities} \label{app:MolOpacs}

See Table \ref{tab:opacities}.

\begin{table}
    \caption{The sources for all considered molecular absorption opacities as included in \textsc{HELIOS} or downloaded from DACE \citep{Grimm:2021}}
    \label{tab:opacities}
    \begin{tabularx}{\columnwidth}{llX}
        \hline
        Species & Opacity Source & Reference \\
        \hline
        \hline
        CH$_4$ & YT34to10 & \begin{tabular}[c]{@{}l@{}}\citet{Yurchenko:2014};\\ \citet{Yurchenko:2017}\end{tabular}\\
        CO$_2$ & UCL-4000 & \citet{Yurchenko:2020}\\
        H$_2$O $^\ast$ & POKAZATEL & \citet{Polyansky:2018}\\
        \hline
        \multicolumn{3}{l}{$^\ast$ H$_2$O only used for Early Mars Model (see Appendix \ref{app:EarlyMars}).}
    \end{tabularx}
\end{table}

\section{Early Mars Benchmark} \label{app:EarlyMars}

To validate our approach and model, we compare it with \citet{Turbet:2020}. They model Early Mars to test whether varying amounts of $\text{H}_2$ or $\text{CH}_4$ can warm the surface sufficiently to support liquid water and compare their own semi-empirically derived $\text{CO}_2$-$\text{H}_2$ and $\text{CO}_2$-$\text{CH}_4$ CIA opacities to the ones from \citet{Wordsworth:2017}.

In their model, they assume 75\,\% of today's solar insolation and a $\text{CO}_2$-dominated atmosphere with varying surface pressures, although we test only the case with $P_\text{surf} = 2$\,bar. The $\text{H}_2\text{O}$ abundance is fixed to 80\,\% relative humidity, and either $\text{H}_2$ or $\text{CH}_4$ is added with a mixing ratio ranging from 0.5 to 10\,\%, constant with pressure.

We use our usual molecular opacities as listed in Appendix A, as well as the default CIA opacity sources for $\text{CO}_2$-$\text{CO}_2$ and $\text{H}_2$-$\text{H}_2$ as marked in Table \ref{tab:cia_overview}. For both CIAs from \citet{Wordsworth:2017}, we used the ones extrapolated from 200 to 150\,K.

\begin{figure}
\begin{center}
    \begin{minipage}{.24\textwidth}
        \includegraphics[width=\textwidth]{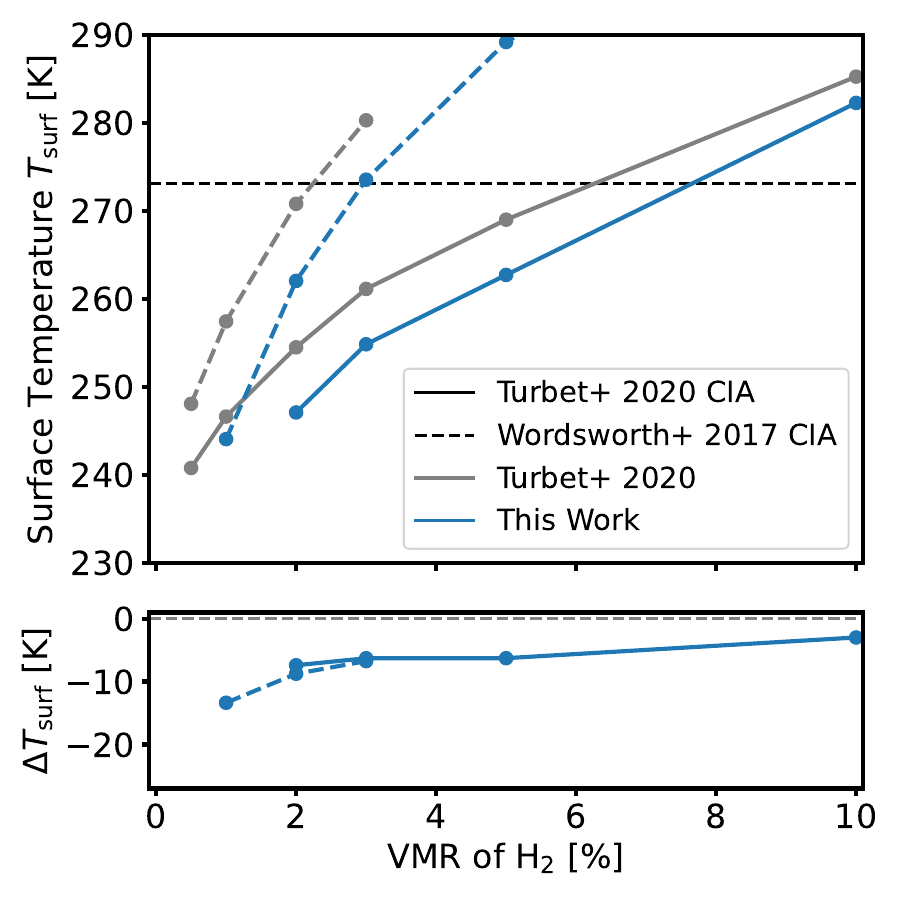}
    \end{minipage}%
    \begin{minipage}{.24\textwidth}
        \includegraphics[width=\textwidth]{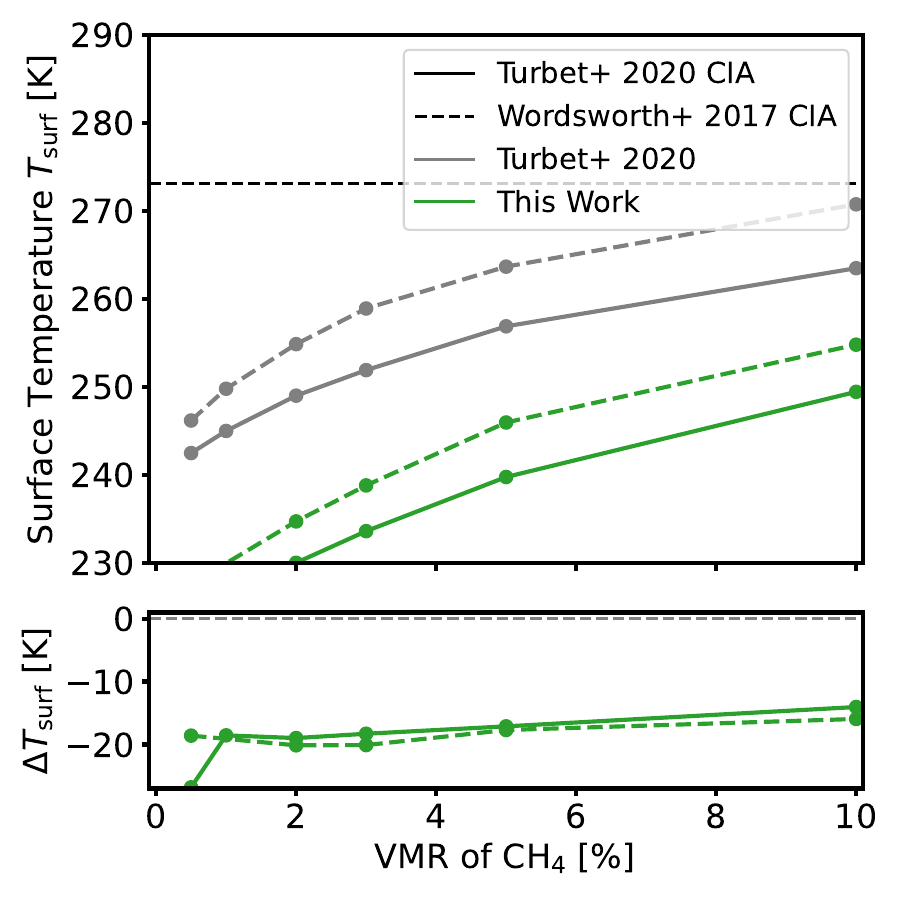}
    \end{minipage}%
    \caption{Comparison of surface temperatures for Early Mars, assuming a 2 bar CO$_2$-dominated atmosphere with varying amounts of either $\text{H}_2$ (left) or $\text{CH}_4$ (right). Solid lines make use of $\text{CO}_2$-$\text{H}_2$ or $\text{CO}_2$-$\text{CH}_4$ CIA data from \citet{Turbet:2020}, while dashed lines use data from \citet{Wordsworth:2017}. Surface temperatures from \citet{Turbet:2020} (grey) are compared with our results (blue or green), with their differences shown in the bottom row. The lower temperatures can be explained by an inaccurate treatment of CO$_2$ pressure broadening.}
    \label{fig:turbet_comparison}
\end{center}
\end{figure}

As shown in Fig. \ref{fig:turbet_comparison}, our calculated surface temperatures are about 14-20\,K lower for the $\text{CH}_4$ case and 3-13\,K lower for the $\text{H}_2$ case. Nevertheless, they show the same trend when switching between CIA sources or increasing the $\text{H}_2$ or $\text{CH}_4$ mixing ratios. The differences could be due to our incorrect pressure broadening of $\text{CO}_2$, similar to \citet{Ozak:2016} who report 15\,K lower surface temperatures, as also noted in \citet{Turbet:2017}.

\bsp	
\label{lastpage}
\end{document}